\documentclass[11pt]{article}

\usepackage[utf8]{inputenc}

\usepackage{fullpage}
\usepackage[margin = 2.5cm]{geometry}

\usepackage{amsmath, amsthm, amssymb}
\usepackage{thmtools}
\usepackage{thm-restate}
\usepackage{mathtools}  
\usepackage{xfrac} 

\usepackage{algorithm}
\usepackage{algpseudocode}

\usepackage{graphicx}
\usepackage{color}

\usepackage[round]{natbib}
\usepackage[hyperindex,breaklinks]{hyperref}
\usepackage{cleveref}
\usepackage{url}

\usepackage{tcolorbox}

\usepackage{tikz}

\usepackage{nicefrac}

\newtheorem{theorem}{Theorem}[section]

\newtheorem{lemma}[theorem]{Lemma}
\newtheorem{definition}[theorem]{Definition}

\DeclareMathOperator{\E}{\mathbf{E}}

\newcommand{\bbR}{\mathbb{R}}

\newcommand{\R}{\mathbb{R}}

\DeclareMathOperator*{\argmax}{arg\,max}

\newcommand{\err}{\mathrm{err}}
\newcommand{\errstar}{\mathrm{err}^*}

\DeclareMathOperator{\REC}{REC}
\DeclareMathOperator{\NRD}{NRD}
\DeclareMathOperator{\FN}{FN}
\DeclareMathOperator{\FP}{FP}

\newcommand{\inner}[2]{\langle #1, #2\rangle}

\title{Error-Tolerant E-Discovery Protocols
\thanks{Equal Contribution. This work is supported by NSF award CCF 1934931 and ECCS 2216970. Thank the anonymous referees for their valuable comments and helpful suggestions. The code is available at \url{https://github.com/dongjs/accountable-tech-assisted-review}}}

\author{Jinshuo Dong\thanks{Northwestern University. Email: \texttt{jinshuo@northwestern.edu}.} \and Jason D. Hartline\thanks{Northwestern University. Email: \texttt{hartline@northwestern.edu}.} \and Liren Shan\thanks{Toyota Technology Institute, Chicago. Email: \texttt{lirenshan@ttic.edu}} \and Aravindan Vijayaraghavan\thanks{Northwestern University. Email: \texttt{aravindv@northwestern.edu}.}}

\begin{document}


\maketitle

\begin{abstract}
    We consider the multi-party classification problem introduced by Dong, Hartline, and Vijayaraghavan (2022) in the context of  
    electronic discovery (e-discovery). 
    Based on a request for production from the requesting party, the responding party is required to provide documents that are responsive to the request except for those that are legally privileged\footnote{The legally privileged documents are protected by privilege, which allows the responding party to resist the mandatory disclosure and withhold such documents in the legal proceeding. Documents that are sensitive or confidential are not necessarily privileged. In this work, we focus on the e-discovery task that aims to find almost all responsive documents among those that are not privileged.}. Our goal is to find a protocol that verifies that the responding party sends almost all responsive documents while minimizing the disclosure of non-responsive documents. 
    We provide protocols in the challenging non-realizable setting, where the instance may not be perfectly separated by a linear classifier. We demonstrate empirically that our protocol successfully manages to find almost all relevant documents, while incurring only a small disclosure of non-responsive documents. We complement this with a theoretical analysis of our protocol in the single-dimensional setting, and other experiments on simulated data which suggest that the non-responsive disclosure incurred by our protocol may be unavoidable.
\end{abstract}

\section{Introduction}

In legal proceedings, a plaintiff (Bob) may need relevant evidence from a corpus of electronic documents possessed by a defendant (Alice).  To acquire this evidence Bob initiates an electronic discovery (e-discovery) process by issuing a request for production to Alice to identify and provide the documents that are responsive to the request.  A few main concerns of the e-discovery process are (a) the efficiency and accuracy by which it is conducted, e.g., \citet{grossman2010technology}; (b) privacy of documents not responsive to the request \citep{stuart2021right}; and (c) accountability for an inaccurate or incomplete review, e.g., via sanctions to Alice's legal team, e.g., \citet{brownvtellermate2014}.

Technology-assisted review (TAR) tools are sometimes used in e-discovery processes to identify relevant documents among the massive collection of electronic documents. \citet{grossman2010technology} show that TAR can accurately retrieve relevant documents with significantly less human effort in manual review.  TAR tools have been approved for use in court cases, e.g., \citet{moorevgroupe2012}.  \citet{kluttz2019automated}, however, argue that accountability is potentially a significant issue for these tools and may limit their adoption; moreover, courts have explicitly not obligated their use, e.g.,  \citet{hylesvnyc2016}. Further discussion of the literature on TAR is in the related work section.

To address the accountability issue of TAR tools, \citet{kluttz2019automated} compares them to medical devices for which there is an approval board that decides which are suitable for use.  They advocate the design of methods and standards for the validation and testing of these tools as the tools of medicine are reviewed by experts before their adoption. They also suggest developing contestable TAR tools that provide transparency, interaction, and configurability.

The starting point for this paper, relative to the above legal scholarship on technology assisted review, is the observation that generally algorithms (which underlie TAR tools) have affordances that other technologies (such as medical devices) do not possess.  Specifically, in certain contexts (a) algorithms can be designed to produce proofs that they are correct in every instance in which they are run (a classical example of these algorithms are zero-knowledge proofs, \citealp{goldwasser1989knowledge}), and (b) algorithms can be partitioned in to multi-party protocols where the computation is divided among the parties in a way that enables them to ensure that information is private (a classical example of these algorithms are secure multi-party computations, \citealp{goldreich1987play}).  A multi-party protocol is one where each party runs an algorithm and these algorithms interact to compute the desired outcome.  

Algorithms are developed from theoretical foundations to practical applications through successive efforts of algorithms researchers. \citet{dong2022classification} demonstrates the plausibility of developing multi-party protocols for e-discovery that satisfy the previous two concerns of proving their correctness (which gives accountability) and maintaining (some) privacy.  They do so in a restricted scenario that is a frequent starting point for developing new machine learning algorithms: the data is represented by points in high-dimensional space, the responsive and non-responsive documents are perfectly separated by a linear function (a.k.a., classifier), and the hand-labeling procedure that Alice or Bob might use to determine which points are responsive and which are non-responsive has perfect accuracy.

\begin{figure*}[t]
    \centering
    \includegraphics[width=0.45\textwidth]{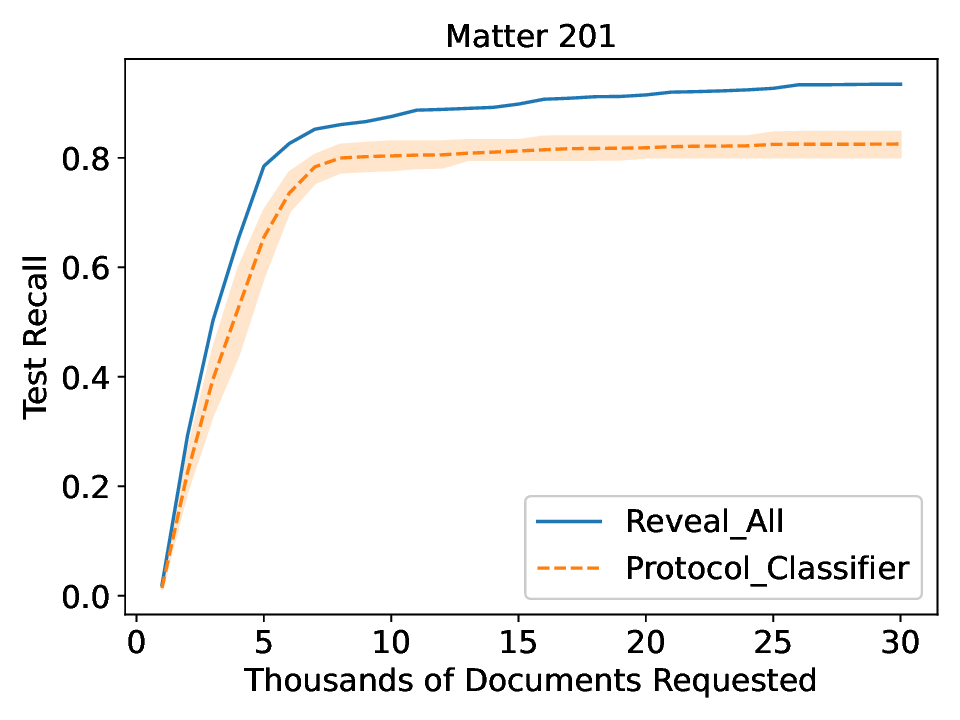}
    \includegraphics[width=0.45\textwidth]{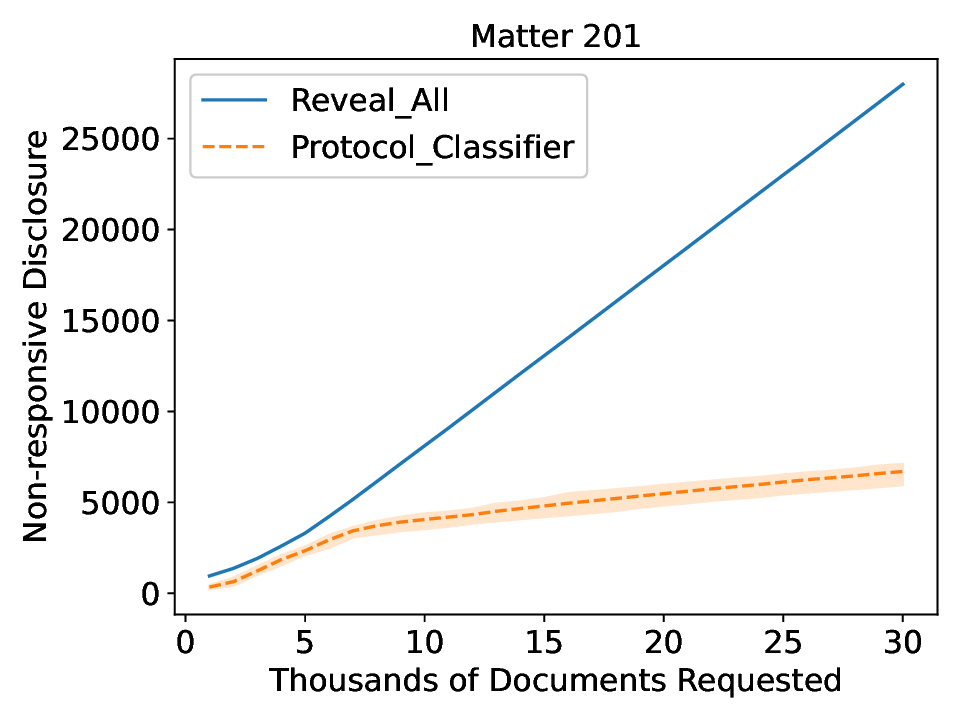}
    \caption{Compared is the Continuous Active Learning (CAL) method of \citet{cormack2014evaluation} with two different internal labeling procedures.  ``Protocol{\textunderscore}Classifier'' uses the labeling protocol that we develop. 
    ``Reveal{\textunderscore}All'' provides the naive baseline of the labeling protocol where all documents that require hand labels are provided to the plaintiff.  
    ``Protocol{\textunderscore}Classifier'' and the baseline ``Reveal{\textunderscore}All'' are both accountable CAL-based protocols, which means they do not rely on the honesty of the defendant. We compare the recall and non-responsive disclosure of these two CAL-based protocols. Note that the version of the CAL method which only reveals responsive documents labeled by hand is not accountable. 
    Experimental results are given for the dataset Matter 201. The left-side figure shows the test recall of two protocols implementing the CAL method that requests $N=1000$ new documents for review in each iteration as the number of iterations $T=1,2,\ldots, 30$. The right-side figure shows the non-responsive disclosure of two protocols as the number of iterations increases.  We repeat our protocol ten times and plot the average recall and the non-responsive disclosure with the corresponding ranges.}
    \label{fig:experiments}
\end{figure*}

Our work relaxes the impractical requirements of \citet{dong2022classification} by giving an e-discovery protocol that finds a good linear classifier when no linear classifier perfectly labels the points and while allowing for the possibility that Alice makes some mistakes in labeling the points.  (These mistakes when caught will be refuted by Bob and resolved in the court.)  \citet{grossman2010technology} and \citet{grossman2012inconsistent} demonstrate that even the best human reviewers may make mistakes in review tasks inadvertently, thus, dealing with mistakes is fundamental to bringing e-discovery protocols closer to practice.
A final property of the protocol we develop is that it fits into the state-of-the-art Continuous Active Learning (CAL) paradigm of \citet{cormack2014evaluation} and thus we can directly compare its performance to standard algorithms for technology assisted review (See \Cref{fig:experiments}).

In classification where there is no classifier that makes no mistakes, known in machine learning as the {\em non-realizable} case, there is a tradeoff between minimizing classification errors that come from false positives and false negatives.  In e-discovery false positives are documents that are incorrectly labeled as responsive when in fact they are non-responsive.  In our analysis we will measure the false-positive rate by tracking this {\em non-responsive disclosure}, i.e., the number of non-responsive documents that are disclosed to Bob.  Notice that our protocols will generally also disclose some documents that it correctly classifies as non-responsive in order to prove that its selected classifier is correct.  These additionally disclosed documents will be included in our analysis of non-responsive disclosure.  False negatives are responsive documents that are incorrectly labeled as non-responsive.  In our analysis we will measure the false-negative rate as the {\em recall}, the fraction of responsive documents that are disclosed by the protocol.  Note for contrast that in the realizable case for the optimal classifier, the non-responsive disclosure is zero and the recall is one.

In this language, \citet{dong2022classification} studied the realizable setting and developed a protocol with recall of one, i.e., all positive documents are disclosed, that provably minimized the non-positive disclosure.  In this case the non-responsive documents disclosed were correctly classified as non-responsive, but must be disclosed to prove that the selected classifier is correct.  This non-responsive disclosure enables the protocol not to reply on the honesty of Alice's labeling of documents.  The TAR algorithms in the literature, e.g., from \citet{cormack2014evaluation}, rely on the honesty of Alice's legal time and, though the classifier they identify may have false negatives, all documents that are actually disclosed to Bob undergo manual review and are labeled as responsive, thus, there is no non-responsive disclosure.  Instead, they aim to tradeoff the number of documents that are reviewed by hand with the recall. There is a trivial method for making these protocols accountable, i.e., so as to not rely on Alice's honesty, which is to release to Bob all documents that are subject to manual review.  In this case, the non-responsive disclosure will be equal to the number of non-responsive documents that are manually reviewed.  We can therefore view existing TAR algorithms as identifying a baseline to which our accountable protocols can be compared.  Our goal in this comparison is to have nearly as large a recall but far lower non-responsive disclosure.

Our contributions are summarized as follows
\begin{enumerate}
	\item We propose a {\em Label-Verification protocol} for the case where the documents are projected into a single-dimension and a linear classifier is simply a threshold above which all points are classified as responsive.  This protocol detects with high probability the manipulation of a dishonest defendant (Alice), while only revealing a small number of non-responsive documents to the plaintiff (Bob). Our e-discovery protocol for the non-realizable setting is constructed by employing the Label-Verification protocol as a subroutine in the Continuous Active Learning (CAL) method proposed by \cite{cormack2014evaluation}. 
    On the one hand, our protocols address the accountability concern in existing TAR methods. On the other hand, our protocols tackle the more challenging non-realizable setting that is not handled by the existing work of~\citet{dong2022classification}.
    We explain our protocols in \Cref{sec:algo}.
 
	\item We demonstrate empirically on review task datasets from the TREC 2009 Legal Track Interactive Task~\citep*{cormack2009machine, hedin2009overview} that compared to the baseline protocol Reveal\_All which is CAL method by revealing all requested documents for accountability, our protocol incurs at most 10\% decrease in recall (i.e., the fraction of discovered responsive documents among all responsive documents) with a significant improvement in terms of the protection of non-responsive documents (in comparison to a naive implementation which is accountable but does not private). Specifically on the above datasets, our protocol has only at most 10\% lower recall, while the non-responsive disclosure is  up to 75\% smaller in comparison to the CAL method where all manually reviewed documents are disclosed. A preview of our results is included in \Cref{fig:experiments}. The details and more results can be found in \Cref{sec:experiment}.

	\item On the theoretical front, we show that the (training) recall of our Label-Verification protocols are provably close to the optimal regardless of the manipulation from the defendant in the single-dimensional case. We also prove that if the defendant reports truthfully, then the non-responsive disclosure of our protocols is small. These results are stated in Theorems~\ref{thm:one-dim-label} and \ref{thm:one-dim-classifier}.
    This gives an explanation for the significant improvement in non-responsive disclosure curves in experiments, as the Label-Verification protocol forms a key subroutine in our e-discovery protocol (See Figure~\ref{fig:experiments}).
    We simulate the defendant with the truthful report in our experiments since it is hard to characterize the best response of the defendant without any assumption on her objective.  
    With proper assumptions on the instance and the objective function of the defendant, we partially characterize the best response of the defendant to our label report protocol in Theorem~\ref{cor:one-dim-label}. We show this best response is close to the truthful report and has similar guarantees.
     Note that in high dimensions, classic results on agnostic learning show that learning a linear classifier in the non-realizable setting is computationally challenging even without any accountability constraints (see the related work section below). 
     Our theoretical results are elaborated in \Cref{sec:theory}.
    
	\item Finally, we show empirically that the non-responsive disclosure is unlikely to have significant room of improvement. We show that such a loss is expected on the simulated data even in the realizable setting, where it is known that the critical point protocol incurs provably minimal non-responsive disclosure~\citep{dong2022classification}. We observe that the fraction of revealed documents in all non-responsive documents are comparable for the two cases. This is made possible by an improved algorithmic implementation of the critical point protocol, which combines a new analysis from the perspective of projective geometry with a classical result from \citet{clarkson1994more}. These results can be found in \Cref{sec:crit}.
\end{enumerate}

\paragraph{Related Work.} This work studies the e-discovery problem with accountability considerations in the linear non-realizable setting. Without any accountability considerations, this is the classic problem of {\em agnostic} learning a halfspace~\citep*{kearns1992toward}. This is a notoriously challenging problem that is believed to be computationally hard in high dimensions see e.g., \cite{kearnsvazirani1994book,guruswami2009hardness, kalai2008agnostically}. Efficient algorithms are known only under additional assumptions e.g., log-concavity of the distribution generating the data points~\citep{awasthi2017localization}, which we do not assume in this work.

In the literature on TAR tools, the work most relevant to our work is the  
continuous active learning (CAL) method proposed by \citet{cormack2014evaluation, cormack2017technology}.
The CAL method iteratively improves the recall in the e-discovery task by interacting with the human reviewers. The CAL method achieves the highest recall in many document retrieval tasks. Therefore, active learning techniques are considered the state-of-the-art in the TAR tools~\citep*{o2015using,zou2020towards}.
Recent advances in natural language processing, especially large language models, provide new techniques for document retrieval tasks. 
\cite{chang2020pre} introduced a two-tower transformer model with pre-training for large-scale retrieval tasks. 
For legal tasks, \citet{louis2021statutory} proposed a RoBERTa-based retrieval model (where RoBERTa is a transformer) on the Belgian Statutory Article
Retrieval Dataset (BSARD). 
In these works, transformers provide more informative embedding of documents, which can potentially be applied in the CAL method and our protocols. We leave this as an interesting open direction.

The multi-party protocols have been widely used to achieve secure computation and preserve privacy in many applications.  \citet{dong2022classification} proposed the e-discovery protocol for the realizable setting. \citet*{goldwasser2021pacverification} studies a different two-party classification problem called PAC verification, that is motivated by delegation of computation. Here the aim is for one party (prover) to convince the other party (verifier) about the correctness of a potential classifier with much fewer samples than it would take for the verifier to learn the classifier. The difference is that in PAC verification, both the prover and the verifier have access to the distribution of labeled examples. In contrast, only one party (defendant Alice) has access to the distribution of labeled examples in our setting.   

\section{Model} 
\label{sec:algo}

In this section, we give a high-level description of our e-discovery protocols. We first define the multi-party protocol model for the e-discovery task and measurements to evaluate such protocols. We then provide our e-discovery protocol based on the continuous active learning (CAL) method introduced by \citet*{cormack2014evaluation}. Despite its efficiency, the CAL method involves manual review of documents and hence is challenging to achieve accountability and privacy simultaneously. We propose the Label-Verification Protocol (\Cref{fig:verification}) to implement the review step accountably with a small disclosure of non-responsive documents. The overall protocol (\Cref{fig:protocol}) then follows by using the Label-Verification Protocol as a subroutine in the CAL method. More details of our protocols can be found in \Cref{sec:experiment}.

\paragraph{Multi-party Protocol Model}
\label{par:multi_party_protocol}

We consider the design of the multi-party protocols as proposed by \citet{dong2022classification}.
There are three parties: Alice (defendant), Bob (plaintiff), and a trusted third party called Trent.
Alice has a set of documents $X$. Bob has issued a request for production, which requires Alice to reveal the documents that are responsive to this request. 
Cloud computing service providers like Gmail and Outlook, or the court system can play the role of Trent. The function $f: X\to \{-1,+1\}$ denotes the true label of these documents regarding this request for production. For each document $x\in X$, it is responsive if $f(x) = 1$; otherwise, it is non-responsive. The words ``label'' and ``responsiveness'' will be used interchangeably in the rest of the paper.
An instance $(X,f)$ is a set of documents $X$ with true labels $f$.
Let $n = |X|$, $n^+ = |\{x\in X: f(x)=1\}|$, and $n^- =  |\{x\in X: f(x)=-1\}|$ be the number of data points, positive points, and negative points in $X$, respectively.

The multi-party protocol contains potentially multiple rounds of interaction between Alice, Bob, and Trent. Both Alice and Bob can label the documents. Trent has access to all documents but can not label them by himself. The outcome $B \subseteq X$ of a multi-party protocol is defined to be a subset of documents in $X$ that are revealed to Bob. 

We now define two measurements that we used to evaluate the multi-party protocols. The goal of the e-discovery protocol is to retrieve nearly all responsive documents reliably while minimizing the privacy loss of the defendant (Alice) caused by revealing non-responsive documents to the plaintiff (Bob). Thus, we measure the efficacy of the e-discovery protocol by the \emph{recall}, a.k.a. true positive rate, which is the most common measurement of TAR tools used in the e-discovery literature. We measure the privacy loss of the defendant in the e-discovery protocol by the \emph{non-responsive disclosure}, which is the number of false positives and negative documents revealed to ensure accountability. Thus, our goal is to design an e-discovery protocol that has a relatively large recall and a small non-responsive disclosure. 

\begin{definition}
    Given an instance $(X,f)$ and an e-discovery protocol $M$, let $B \subseteq X$ be the outcome of this protocol, which is the set of documents revealed to the plaintiff. The \emph{recall} of this outcome is the fraction of responsive documents that are retrieved in $B$
    \begin{equation}
        \REC(B) = \frac{|\{x \in X: x \in B, f(x) = +1\}|}{|\{x \in X: f(x) = +1\}|}.
    \end{equation}
    The \emph{non-responsive disclosure} of this outcome $B$ is the number of non-responsive documents revealed to the plaintiff
    \begin{equation}
        \NRD(B) = |\{x \in X : x \in B, f(x) = -1\}|.
    \end{equation}
\end{definition}

We now define the non-realizable setting and the benchmark for e-discovery protocols. As a natural language processing tool, the e-discovery protocol will first embed all documents into an Euclidean space with dimension $d$. With an abuse of notation, let $X = \{x_1,\dots,x_n\} \subset \bbR^d$ be the embedding of all documents. For any two vectors $x,y \in \bbR^d$, we use $x \cdot y$ to denote the inner product of these two vectors. We consider the linear classification on this embedded instance $(X,f)$. A linear classifier $h : X \to \{-1,+1\}$ partitions the $d$-dimensional space into two parts with a hyperplane $w\cdot x + b = 0$ where $w \in \bbR^d$ and $b \in \bbR$. The vector $w$ is the normal vector of this hyperplane. For each document $x\in X$, $h(x) = 1$ if $w\cdot x + b \geq 0$; otherwise, $h(x) = -1$. The error of a linear classifier $h$ on the embedded instance $(X,f)$ is defined to be the number of misclassified documents
$\err(h) = |\{x \in X: h(x) \neq f(x)\}|$. Let $h^*$ be the optimal linear classifier for the instance $(X,f)$ with the minimum error $\err^* = \err(h^*)$. 

An embedded instance $(X,f)$ is realizable (or linearly separable) if and only if the optimal classifier $h^*$ perfectly classifies all documents, i.e. $\err^* = 0$. Otherwise, this instance $(X,f)$ is called non-realizable, which implies the error of the optimal linear classifier is strictly greater than $0$. 
Let $\FN(h) = |\{x \in X: f(x) = 1, h(x) = -1\}|$ be the false negative error of a linear classifier $h$. Let $\FN^*= \FN(h^*)$ be the false negatives of the optimal linear classifier.
In the non-realizable setting, revealing $B^* = \{x\in X : h^*(x) = 1\}$ according to the optimal linear classifier achieves a recall $\REC^*=\REC(B^*) = 1-\FN^*/n^+ < 1$. Thus, the minimum error $\err^*$ and this recall $\REC^*$ given by $h^*$ reflect the inherent difficulty of this instance. We use the optimal linear classifier and its recall $\REC^*$ as our benchmark and aim to design a protocol that achieves a recall close to $\REC^*$ with a small non-responsive disclosure.

\paragraph{The CAL Method} 
\label{par:basic_settings_and_the_cal_method}


To find responsive documents efficiently, \citet*{cormack2014evaluation} proposed a technology-assisted review (TAR) method, continuous active learning (CAL). 
The CAL method first embeds all documents into a high-dimensional Euclidean space $\bbR^d$. 
The CAL method iteratively maintains and grows a subset of documents $S\subset X$ (with random initialization) which are labeled by human reviewers. In each iteration, the classifier is first updated by training a linear support vector machine (SVM) on the current $S$. 
Each remaining document is assigned a score based on the SVM classifier, which is the projection to the normal vector $w$ of the hyperplane corresponding to the SVM classifier.
Then the $N$ documents with the highest scores among the remaining documents are actively chosen and manually reviewed to confirm their labels. The score of each of the remaining documents represents how likely this document is to be labeled responsive according to the updated classifier. Thus, these $N$ documents are the most likely to be labeled responsive in the remaining unreviewed documents based on the updated classifier. After the manual review, these $N$ labeled documents are added to $S$ and this finishes one iteration.
The pseudo-code is summarized in Algorithm~\ref{alg:CAL}.

\begin{algorithm}
\caption{Continuous Active Learning (CAL)}\label{alg:CAL}
    \begin{algorithmic}[1]
        \State \textbf{Input:} a collection of documents $X$, a document review query, parameters $T$ and $N$, where $T$ is the number of iterations and $N$ is the number of requested documents in each iteration.
        \State \textbf{Output:} a set of responsive documents
        \State Let set $S=\varnothing$.
        \State Select $N$ documents randomly from the results of a seed query search.
        \For{$i = 1,2, \cdots, T$}
            \State {\em Manually review} $N$ new documents and add them with their labels to set $S$.
            \State Train a linear classifier with the support vector machine on dataset $S$.
            \State Compute the score of each remaining document as the projection to the normal vector $w$ of the hyperplane corresponding to the current linear classifier. 
            \State Select $N$ documents from $X\setminus S$ with the highest scores.
        \EndFor
        
        \State \Return responsive documents in $S$. 
    \end{algorithmic}
\end{algorithm}

\paragraph{E-discovery Protocol} 
\label{par:E-discovery protocol}
We now describe our e-discovery protocol, which implements the CAL method reliably with a verification process. 
The CAL method is an e-discovery algorithm that makes repeated calls to a primitive e-discovery algorithm (which is the manual review in the original CAL).  When that primitive e-discovery algorithm is a multi-party protocol (as defined previously), then the combination is a multi-party protocol as well.

Note that the CAL method requires the manual review of new documents added in each iteration (Step 9 of Algorithm~\ref{alg:CAL}), which causes accountability concerns. Alice may intentionally hide responsive documents in this manual review process. 
To implement this step reliably and accountably, 
in our protocol, Trent simulates the CAL method by calling on Alice and Bob in the Label-Verification protocol. The framework of our e-discovery protocol is shown in Figure~\ref{fig:protocol}. 
In each iteration, Trent computes a set of $N$ new documents as in Step 9 of Algorithm~\ref{alg:CAL}. Let $R$ be this set of documents embedded in the single-dimensional space $\bbR$ according to their scores computed in Step 8 of Algorithm~\ref{alg:CAL}. Then, Trent call the Label-Verification protocol to retrieve responsive documents in this sub-instance $(R,f)$.



\begin{figure}[tb]
    \centering
    \begin{tcolorbox}
        \begin{enumerate}
            \item Input the set of all documents $X$ and parameters $T, N$ of the CAL method.
            \item Trent simulates the CAL method on $X$ with $T$ iterations.
            \item In each iteration $i \leq [T]$, let $R \subseteq X \setminus S$ be the set of $N$ new documents selected in Step 9 of Algorithm~\ref{alg:CAL}.
            \item Trent runs the Label-Verification protocol with Alice and Bob on the sub-instance $(R,f)$, where $f$ is the true label function and documents in $R$ are embedded in $\bbR$ with their scores computed in Step 8 of Algorithm~\ref{alg:CAL}. 
            \item Trent adds these documents $R$ and their labels returned by the Label-Verification protocol to set $S$ and start a new iteration of the CAL method. 
            \item Output the responsive documents in $S$, which are revealed to Bob.
        \end{enumerate}
    \end{tcolorbox}
    \caption{E-discovery Protocol}
    \label{fig:protocol}
\end{figure}

\paragraph{Label-Verification Protocol} 
\label{par:label_verification_protocol}
We now describe the main subroutine of our protocol, the single-dimensional Label-Verification protocol. The Label-Verification protocol is a multi-party protocol, which aims to retrieve responsive documents in the input instance $(R,f)$ accountably.  We first introduce a high-level, ``template'' protocol, which covers protocols for non-realizable data proposed in this work. We then give the specific implementations.

\begin{figure}[ht]
    \centering
    \begin{tcolorbox}
        \begin{enumerate}
            \item Input a set of documents $R$ embedded in the single-dimensional space $\bbR$
            \item Trent requests Alice to report the labels of documents in $R$.
            \item Trent iteratively selects documents in $R$ and reveals those documents to Bob.
            \item Bob labels received documents and reports the labels to Trent.
            \item Trent checks the agreement of reports from Alice and Bob. If Bob disagrees with Alice's label for a document, then this document is sent to the court to settle. 
            \item Trent checks the stopping condition and decides to stop or go to Step 2 (send more documents to Bob).
            \item Output the labels of all documents in $R$ as follows. Each document is labeled by the court decision if this document is sent to the court; otherwise, it is labeled by Alice's report.
        \end{enumerate}  
    \end{tcolorbox}
    \caption{Label-Verification Protocol}
    \label{fig:verification}
\end{figure}

The template label-verification protocol we propose is the following iterative process. It first requests Alice to report the labels of all documents in $R$. Then, it iteratively sends some documents in $R$ to Bob for verification until the stopping condition is satisfied.  
If Bob disagrees with Alice on the label of a document, then this document is sent to the court to settle. If the court disagrees with Alice on the label of this document, then we say Alice made an error on this document. The template Label-Verification protocol is shown in Figure~\ref{fig:verification}.

There is a trivial Label-Verification protocol, which reveals all documents in $R$ to Bob for verification in Step 3 of Figure~\ref{fig:verification}. 
This trivial protocol has perfect accountability (recall) on the hand-labeled documents $R$ since Bob receives all responsive documents in $R$. While this protocol also reveals all non-responsive documents in $R$ to Bob. Thus, it has the maximum non-responsive disclosure on the sub-instance $(R,f)$. We use the CAL with this trivial protocol as the baseline, denoted as Reveal\_All. 

Two details need to be specified in order to recover the concrete protocols in our work: (1) how Trent selects the subset of documents in Step 3; and (2) the stopping condition in Step 6. In the non-realizable case, the following specifications yield our two Label-Verification protocols. Further details can be found in \Cref{sec:theory} 
\begin{itemize}
    \item Selection: Trent first computes the optimal classifier threshold $t_A \in \bbR$ for Alice's report. Then, Trent sends all documents in $[t_A,\infty)$ to Bob for verification. If $t_A$ is the optimal classifier for the true label, then the recall is already the recall $\REC^*$ of the optimal classifier.
    If $t_A$ is not the optimal classifier for the true label, then Alice hides positive points close to $t_A$ as negative points.
    To detect this case, points in $(\infty,t_A]$ reported as negative by Alice are sampled independently in the decreasing order of their positions. Generally, the points closer to $t_A$ are sampled earlier with higher probabilities. The specific probability depends on the honesty of Alice's prior reports. These sampled points are revealed to Bob.
    \item Stopping condition: The Label-Verification protocol stops if enough errors are detected in Alice's report, or when no more documents can be sampled (If a document is not sampled in the first time, then this document can not be sampled again.). In Step 5, this document is sent to the court to settle if Bob disagrees with Alice's label for this document. If a document is sent to the court, then the protocol outputs the court decision as the label of this document. Otherwise, it labels this document by Alice's report. Note that all points labeled as positive in the output are already verified by both Alice and Bob. We say Alice made an error on the label of a document if and only if the court disagrees with Alice's label.
\end{itemize}

\section{Empirical Results}
\label{sec:experiment}

In this section, we show the empirical results of our protocols. Our protocols are constructed by combining our Label-Verification protocols in Section~\ref{sec:theory} with the CAL method shown in Algorithm~\ref{alg:CAL}.

We use the review task, Matters 201 and 202 from the TREC 2009 Legal Track Interactive Task~\citep*{cormack2009machine, hedin2009overview}.
Matter 201 is the requests for production of all documents related to the Company's engagement in prepay transactions.
Matter 202 is the requests for production of all documents related to the Company's engagement in transactions characterized as compliant with FAS 140 or FAS 125. 
The information of datasets is provided in Table~\ref{tab:dataset}, including the number of total documents and the number of responsive documents for each review task. 
We refer to~\citet{cormack2014evaluation} for a more detailed description of these review tasks.

\begin{table}[H]
\begin{center}
\begin{tabular}{ |c|c|c| } 
 \hline
 Matter & Number of Documents & Number of Responsives \\
 \hline
 201 & 723537 & 2154 \\ 
 202 & 723537 & 8734 \\ 
 \hline
\end{tabular}
\end{center}
\caption{Information of datasets, Matters 201 and 202 from the TREC 2009 Legal Track Interactive Task.}
\label{tab:dataset}
\end{table}


\begin{figure*}[ht]
    \centering
    \includegraphics[width=0.45\textwidth]{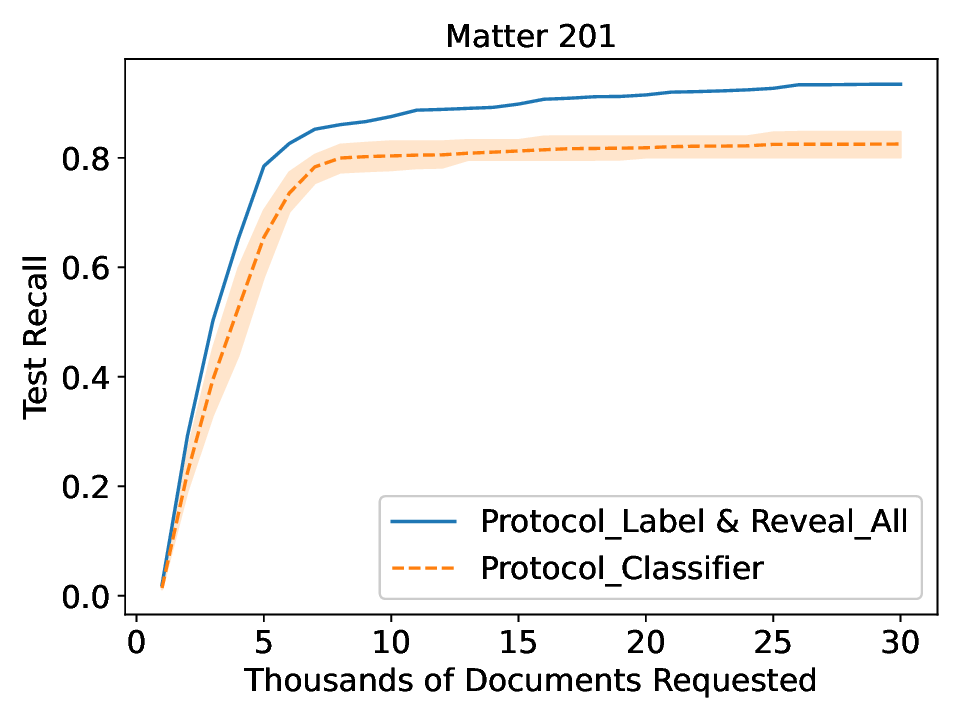}
    \includegraphics[width=0.45\textwidth]{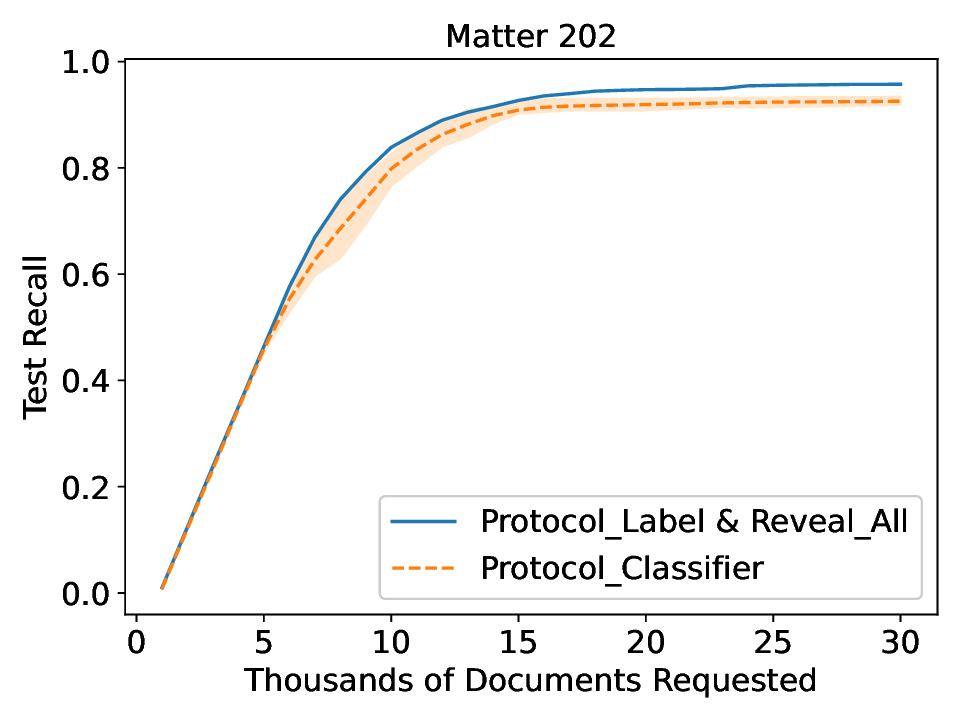}
    \caption{Test recall of Protocol\_Label (CAL with the sampling protocol for label report), Protocol\_Classifier (CAL with the sampling protocol for classifier report) and Reveal All (CAL by revealing all requested documents) on the datasets Matter 201 and 202. We plot the test recall of three protocols implementing the CAL method with the number of iterations $T=1,2,\cdots, 30$ and select $N=1000$ new documents for review in each iteration. We repeat our protocols for $10$ times and plot the average recall with the corresponding ranges.}
    \label{fig:recall}
\end{figure*}

\begin{figure*}[ht]
    \centering
    \includegraphics[width=0.45\textwidth]{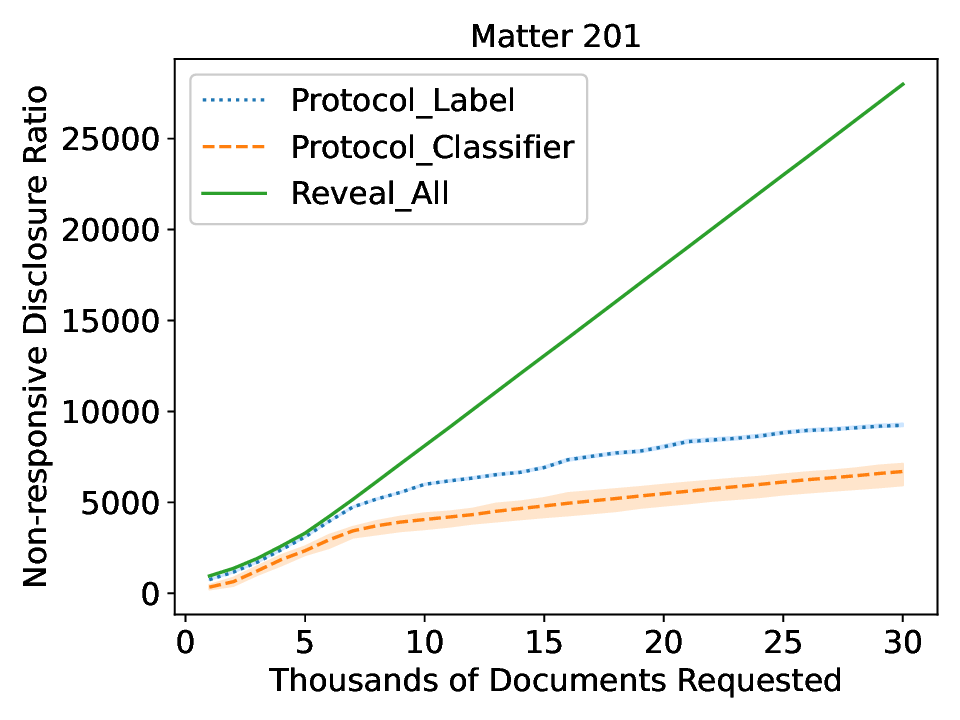}
    \includegraphics[width=0.45\textwidth]
    {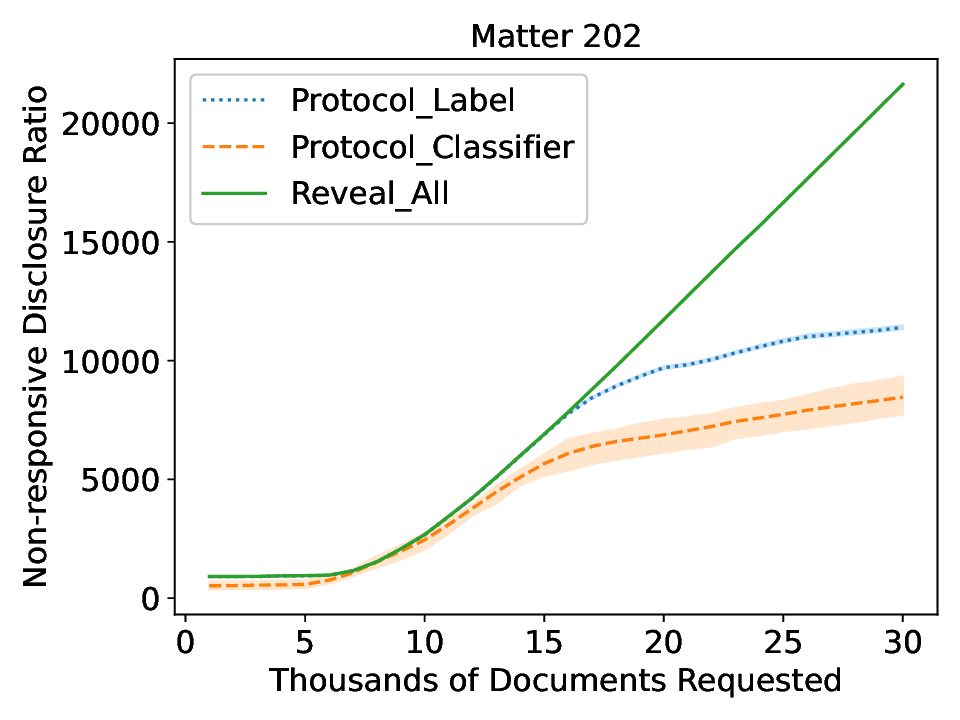}
    \caption{Non-responsive disclosure of Protocol\_Label (CAL with the sampling protocol for label report), Protocol\_Classifier (CAL with the sampling protocol for classifier report) and Reveal All (CAL by revealing all requested documents) on the datasets Matter 201 and 202. We plot the non-responsive disclosure of three protocols with the number of iterations $T=1,2,\cdots, 30$ in the CAL method that selects $N=1000$ new documents for review in each iteration. We repeat our protocols for $10$ times and plot the average non-responsive disclosure with the corresponding ranges.}
    \label{fig:privacy}
\end{figure*}

Our protocols implement the CAL method (Algorithm~\ref{alg:CAL}) with parameters $N=1000$ and $T=1,2,\cdots, 30$. 
The CAL method is used as a TAR tool to find responsive documents by iteratively selecting new documents for manual review. 
Note that this method requires labeling the $1000$ documents that are selected in each iteration. 
We consider the multi-party implementation of the CAL method. Trent implements the CAL method by using a Label-Verification protocol to label documents in each iteration. 
In the experiments, we implement the following three protocols:
\begin{itemize}
    \item Reveal\_All: A naive Label-Verification protocol is to reveal all $1000$ documents to Bob for verification.  We use Reveal\_All to denote the e-discovery protocol by combining the CAL method with this naive protocol. This protocol has the same recall as the CAL method, but might incur a large non-responsive disclosure. We use this protocol as our baseline. 
    \item 
    Protocol\_Label: We use the sampling protocol for label report in Section~\ref{sec:label_report} (Algorithm~\ref{alg:one_dim_label}) as the Label-Verification protocol. Protocol\_Label is the e-discovery protocol by combining the CAL method with this Label-Verification protocol. We set parameters $\delta = 0.01$ and $k=1$ in the sampling protocol for label report.
    We simulate Protocol\_Label by assuming Alice truthfully reports all true labels of documents. 
    \item 
    Protocol\_Classifier: We use the sampling protocol for classifier report in Section~\ref{sec:classifier_report} (Algorithm~\ref{alg:one_dim_classifier}) as the Label-Verification protocol. Protocol\_Classifier is the e-discovery protocol by combining the CAL method with this Label-Verification protocol. We set parameter $\delta = 0.01$ in the sampling protocol for classifier report. We simulate Protocol\_Classifier by assuming Alice truthfully reports the optimal classifier.
\end{itemize}


\paragraph{Results} 
We now describe our empirical results for recall and non-responsive disclosure of these three protocols. 
Figure~\ref{fig:recall} shows the test recall of three protocols as the number of documents that Trent requests Alice to review increases. Since these three protocols are based on the CAL method, the number of hand-labeled documents increases by $N=1000$ for each iteration as the number of iterations used in the CAL method increases. 
Since the baseline Reveal\_All reveals all hand-labeled documents to Bob for verification, it achieves full accountability on the label of hand-labeled documents. 
Thus, Reveal\_All achieves the same recall as the CAL method with true labels of hand-labeled documents.

Protocol\_Label uses a strict Label-Verification protocol that requires Alice to report the labels of all hand-labeled documents correctly.
For Protocol\_Label, we assume that Alice truthfully reports all true labels of hand-labeled documents. With this assumption, Protocol\_Label has the same labels of the hand-labeled documents in each iteration as Reveal\_All. Thus, the recall of Protocol\_Label is the same as that of the baseline Reveal\_All. 
In Section~\ref{sec:theory}, we show that Protocol\_Label guarantees a small (training recall) loss in each iteration, and Alice would be partially truthful under proper assumptions. We expect that Protocol\_Label will have a small test recall loss compared to the baseline in practice. 

Protocol\_Classifier further relaxes the requirement on the report of Alice. It uses a Label-Verification protocol that only requires Alice to report the optimal threshold on the hand-labeled documents.
Figure~\ref{fig:recall} shows that when Alice truthfully reports the optimal threshold, the recall of Protocol\_Classifier decreases by at most 10\% compared to the recall of Reveal\_All. 

Figure~\ref{fig:privacy} shows the non-responsive disclosure of three protocols as the number of hand-labeled documents increases. 
The baseline Reveal\_All reveals all hand-labeled documents to Bob for verification. Thus, it has the maximum non-responsive disclosure, which equals the number of non-responsive documents in all hand-labeled documents.

Instead of revealing all hand-labeled documents, Protocol\_Label uses the Label-Verification protocol for label report (Algorithm~\ref{alg:one_dim_label}) to sample a subset of requested documents. The non-responsive disclosure of Protocol\_Label is up to 50\% lower than that of Reveal\_All. 

Protocol\_Classifier is less strict on Alice's report and is tolerant of errors in Alice's report. It uses a Label-Verification protocol for classifier report (Algorithm~\ref{alg:one_dim_classifier}) with small and adaptive sampling probabilities. Protocol\_Classifier further decreases the non-responsive disclosure by up to 20\% compared to that of Protocol\_Label.

According to empirical results, when Alice truthfully reports all labels of requested documents, Protocol\_Label achieves the same recall as the baseline with a much smaller non-responsive disclosure than the baseline. With the analysis in Section~\ref{sec:theory}, we expect that in practice, Protocol\_Label will have a small recall loss compared to the baseline with a small non-responsive disclosure. Protocol\_Classifier is tolerant of errors in Alice's report and only requires Alice to report the optimal threshold. The empirical results show that when Alice truthfully reports the optimal threshold, Protocol\_Classifier has at most 10\% recall loss compared to the baseline with less than $1/3$ of the non-responsive disclosure of the baseline.

\section{Theoretical Guarantees}
\label{sec:theory}


In this section, we describe the Label-Verification protocols used in our e-discovery protocols for the non-realizable setting. 
We prove theoretical guarantees of recall and non-responsive disclosure for these protocols. These guarantees are for each single call of the Label-Verification protocol. We do not have theoretical results for the combination of these protocols with the CAL method.

We consider the one-dimensional instance $(R,f)$ as the input of our Label-Verification protocols, where the set of documents are embedded in the one-dimensional space, $R = \{x_1,x_2,\dots,x_N\} \subset \bbR$. Note that for high-dimensional instances, our e-discovery protocols Protocol\_Label and Protocol\_Classifier simulate the CAL method and iteratively call these Label-Verification protocols on a subset $R \subset X$ with the highest scores according to the current linear classifier. Thus, this subset of documents $R$ and their scores given by the current linear classifier form a one-dimensional sub-instance. Our analysis provides guarantees for each call of the Label-Verification protocol on the sub-instance $(R,f)$.

We now define some useful notions used in our Label-Verification protocol for label report (Algorithm~\ref{alg:one_dim_label}) and Label-Verification protocol for classifier report (Algorithm~\ref{alg:one_dim_classifier}). 
For each document $x \in R$, let $f(x) \in \{-1,1\}$ be the true label of this document $x$, and $f_A(x)$ be the label of this document $x$ reported by Alice. 
We use $t^*$ to denote the optimal threshold to classify these points with true labels, which minimizes the error 
$$
t^* \in \arg\min_{t\in \bbR} |\{x\in R: f(x) \neq \text{sign}(x-t)\}|.
$$ 
If there is more than one optimal threshold, we let $t^*$ be the largest one. 
Similarly, we use $t^*_A$ to denote the optimal threshold to classify the data points with Alice's labels, i.e.
$$
t^*_A \in \arg\min_{t \in \bbR} |\{x\in R: f_A(x) \neq \text{sign}(x-t)\}|.
$$ 
If there are multiple optimal thresholds for Alice's report, then we let $t^*_A$ be the smallest one. 
Let $\err(t) = |\{x\in R : f(x) \neq \text{sign}(x-t)\}|$ be the error of the threshold $t$ on the true labels. Let $\err_A(t) = |\{x\in R: f_A(x) \neq \text{sign}(x-t)\}|$ be the error of the threshold $t$ on Alice's report. For simplicity, we use $t^*=t^*(R,f)$ and $\errstar = \err(t^*)$ to denote the best threshold and the optimal error. 

The instance $(R,f)$ is non-realizable, which means responsive and non-responsive documents in $(R,f)$ can not be perfectly separated by a linear classifier, i.e. the error of the best classifier $\err(t^*)$ is strictly greater than $0$. We consider the two different settings for Alice's report in the Label-Verification protocol. In Section~\ref{sec:label_report}, we consider the Label-Verification protocol in which Alice is requested to report all labels of documents in $R$ correctly. In Section~\ref{sec:classifier_report}, we provide a robust Label-Verification protocol in which Alice is only requested to report the best classifier on $(R,f)$.

\subsection{Report Labels}\label{sec:label_report}
We first consider the setting in which Alice is requested to report all labels of documents in $R$ correctly. We provide a Label-Verification protocol for this strict label report setting. We show the following theorem for this protocol. The proofs in this section are deferred to Appendix~\ref{sub:proofs_in_sec:label_report}.

\begin{restatable}{theorem}{thmlabel}
\label{thm:one-dim-label}
    Given an one-dimensional instance $(R,f)$, failure probability $\delta \in (0,1)$ and error tolerance $k \geq 1$, Label-Verification Protocol for Label Report (Algorithm~\ref{alg:one_dim_label}) satisfies
    \begin{enumerate}
        \item (Recall) The recall is at least $1-(\err^* +k-1)/N^+$ with probability at least $1-\delta$;
        \item (Non-responsive disclosure) If Alice reports the true labels $f$ on $R$, then the expected non-responsive disclosure is at most $$\left(2+\frac{2\,\errstar}{k}\right) \ln N^- \ln(1/\delta) + \err^*,$$
    \end{enumerate}
    where $N$ is the number of data points in $R$, $N^+, N^-$ are the number of positive points and negative points in $R$ respectively, $\errstar$ is the optimal error. 
\end{restatable}

Note that $\err^* = \FN^* + \FP^*$, where $\FN^*$ and $\FP^*$ are the number of false negatives and false positives given by the optimal classifier on $(R,f)$. 
The recall of the optimal classifier is $\REC^* = 1-\FN^*/N^+$.  
For $k=1$, Label-Verification Protocol for Label Report is guaranteed to achieve $1-\err^*/N^+ = \REC^* - \FP^*/N^+$ recall on $(R,f)$ with probability as least $1-\delta$. 
This means, except for the false negatives in the optimal classifier, Alice can potentially hide $\FP^*$ extra positives as negatives. 
Alice can utilize the false positives of the optimal classifier to hide the same number of positive points as negatives.
The error tolerance $k \geq 1$ captures the trade-off between the recall and non-responsive disclosure. For smaller $k$, this protocol has a stronger recall guarantee with a larger non-responsive disclosure for verification.

Consider a nice instance $(R,f)$ with a constant optimal error $\err^*$. With a small constant error tolerance $k$, this protocol has a small recall loss compared to the recall of the optimal classifier $\REC^*$. With a small constant failure probability $\delta$, this protocol has $O(\log N^-)$ non-responsive disclosure if Alice reports true labels, which is much smaller than the number of true negatives in $R$. 



This Label-Verification protocol works as follows. It first computes the optimal classifier $t^*_A$ for Alice's report $f_A$. Then, it asks Bob to verify all documents reported as positive by Alice and documents classified as positive by $t^*_A$. It samples the rest documents independently in decreasing order of their positions. The document with the $i$-th highest position is sampled and sent to Bob with probability proportional to $1/i$. 
If Bob disagrees with Alice on the label of this document, then this document is sent to the court to settle. The sampling process stops if it goes through all documents reported as negative by Alice or if the protocol detects Alice made an error. If the protocol detects Alice made an error, then Trent sends all documents reported as negative by Alice to Bob to verify and sends any disagreement to the court to settle.
Finally, the protocol outputs the label of each document as the court decision if this document is sent to the court; otherwise outputs the label of this document reported by Alice.
In this output, all documents labeled as positive are either verified by Bob or settled by the court. Thus, documents labeled as positive in the output are always correct.  
The pseudo-code of our sampling protocol is shown in Algorithm~\ref{alg:one_dim_label}.

\begin{algorithm}[ht]
\caption{Label-Verification Protocol for Label Report}\label{alg:one_dim_label}
    \begin{algorithmic}[1]
        \State \textbf{Input:} a set of documents $R$ embedded in the one-dimensional space $\bbR$, error tolerance $k \geq 1$, and failure probability $\delta \in (0,1)$
        \State \textbf{Output:} labels of documents in $R$ 
        \State Alice sends all points and their labels $\{(x,f_A(x))\}_{x\in X}$ to Trent.
        \State Trent computes the optimal threshold $t^*_A$ for all points with labels reported by Alice.
        \State Trent sends all positive points reported by Alice and all points $x \in (t^*_A, +\infty)$ to Bob. 
        \State Trent goes through all points in $(-\infty,t^*_A)$ labeled as negative by Alice in decreasing order of their embedding, $t^*_A > x_1 > x_2 > \cdots > x_m$.
        \For{each negative point $x_i$}
            \State Trent sends this point to Bob with probability $p_i = \min\{1, c/i\}$, where 
            $$c = \left(2+\frac{2\,\err_A(t^*_A)}{k}\right) \ln(1/\delta).$$
            \State Bob labels this point and sends the label to Trent.
            \State If Bob labels this point as positive, then Trent sends this point to the court. If this point is a true positive, then Trent sends all points to Bob. Bob sends the labels of these points to Trent. Trent sends the points that have different labels from Alice and Bob to the court to settle. 
        \EndFor
        \State Output the label of each document $x \in R$ as $f_A(x)$ if $x$ is not sent to the court; otherwise, output the label of this document decided by the court.
    \end{algorithmic}
\end{algorithm}

Note that the non-responsive disclosure bound in Theorem~\ref{thm:one-dim-label} applies when Alice reports the true labels $f$. We now examine the best response of Alice to this protocol. To do this, we first specify the loss function of Alice $L_A$. Note that Alice will benefit from the responsive documents that are not retrieved and sent to Bob. Alice also has a privacy loss due to the disclosure of any non-responsive documents to Bob. Thus, Alice has a loss function $L_A$ on the outcome $B \subset R$ of the protocol that depends on the recall $\REC(B)$ and the non-responsive disclosure $\NRD(B)$. We assume the loss function of Alice is linear, $L_A(B) = \NRD(B) + c \cdot N^+ \cdot (\REC(B)-1)$.

By assuming Alice's loss function $L_A$ is linear and the instance $(R,f)$ has enough number of true negatives, we have the following theorem on the best response of Alice. The proof is deferred to Appendix~\ref{sub:proofs_in_sec:label_report}.

\begin{restatable}{theorem}{corlabel}
\label{cor:one-dim-label}
    Suppose Alice has a linear loss function on the outcome $B \subset R$, $L_A(B) = \NRD(B) + c \cdot N^+ \cdot (\REC(B)-1)$ for some constant $c\geq 0$. For any $\delta \in (0,1/3)$ and $k \geq 1$, consider an instance $(R,f)$ with the number of negative points at least 
    $$
    N^- > \max\{3(2+2\errstar/k)\cdot\ln(1/\delta)\cdot\ln N^- + 3\err^*, c \cdot N^+\}.
    $$
    Alice's best response to Label-Verification Protocol with Label Report satisfies that
    \begin{enumerate}
        \item (Recall) The recall is always at least $1-(\err^*+k)/N^+$;
        \item (Non-responsive disclosure) The expected non-responsive disclosure is at most $$(2+2\errstar/k)\ln N^- \ln(1/\delta) + (c+1)\err^*+ck.$$
    \end{enumerate} 
    
    Particularly, for $k = 1$, Alice's best response truthfully reports the labels of all points classified as positive by the optimal classifier of $(R,f)$, which provides the recall at least $\REC^*$. 
\end{restatable}

If Alice does not report the true labels, then the outcome of our protocol can be divided into two cases:
(1) If the protocol detects a positive point hidden by Alice, then it reveals all documents to Bob for verification. In this case, the recall is $1$ and non-responsive disclosure is the number of true negatives $N^-$. (2) If the protocol does not detect any hidden positive, then the bounds for recall and non-responsive disclosure in Theorem~\ref{thm:one-dim-label} hold. 
By assuming enough true negatives $N^-$ and Alice's loss function, the loss in the first case is larger than that of the second case. Then, Alice will report the true labels for a large portion of documents in the best response. However, if some false positives of the optimal classifier are far away from the classifier boundary, which are outliers, then Alice may hide these positives as negative with a very small probability of being detected. Hence, we only partially characterize Alice's best response. 

\subsection{Report a Classifier}\label{sec:classifier_report}
We now consider a setting where Alice is only requested to report a good linear classifier on $R$ instead of labeling all documents correctly.
This setting is tolerant of errors made by Alice since Alice is requested to find a good linear classifier and is allowed to make errors. 
We provide an error-tolerant Label-Verification protocol for this classifier report setting. 
We show the following theorem for this Label-Verification protocol (Algorithm~\ref{alg:one_dim_classifier}). The proof of this theorem is in Appendix~\ref{sub:proofs_in_sec:classifier_report}

\begin{restatable}{theorem}{thmclassifier}
\label{thm:one-dim-classifier}
    Given an one dimensional instance $(R,f)$ and failure probability $\delta \in (0,1)$, Label-Verification Protocol for Classifier Report (Algorithm~\ref{alg:one_dim_classifier}) satisfies
    \begin{enumerate}
        \item (Recall) The recall is at least $1-3\err^*/N^+$ with probability at least $1-\delta$;
        \item (Non-responsive disclosure) If Alice reports an optimal classifier for $(R,f)$, then the expected non-responsive disclosure is at most 
        $$2\,\err^* \ln N \ln(N/\delta) + \err^*,$$
    \end{enumerate}
    where $N$ is the number of data points, $N^+$ is the number of positive points, and $\err^*$ is the optimal error for this instance. 
\end{restatable}

\begin{algorithm}[tb]
\caption{Label-Verification Protocol for Classifier Report}\label{alg:one_dim_classifier}
    \begin{algorithmic}[1]
        \State \textbf{Input:} a set of $N$ documents $R$ embedded in the one-dimensional space $\bbR$ and failure probability $\delta \in (0,1)$
        \State \textbf{Output:} labels of documents in $R$ 
        \State Alice sends all points $X$ and a classifier $t_A$ to Trent.
        \State Trent asks Alice to label all points in $[t_A, +\infty)$. Trent sends all points in $[t_A, +\infty)$ to Bob. Bob sends the labels of these points to Trent. Trent sends the points that have different labels from Alice and Bob to the court to settle. 
        \State Trent goes through all points in $(-\infty,t_A)$ in decreasing order, $t_A > x_1 > x_2 > \cdots > x_m$.
        \State Set the counts of positives and negatives $M^+ = M^- = 0$ and the weight $W = 1$.
        \For{each point $x_i$}
            \State Trent sends this point to Bob with probability $p_i = \min\{1, c/W\}$, where 
            $$c = 2\ln(N/\delta).$$
            \State Bob labels this point and sends the label to Trent.
            \State If Trent does not send this point to Bob or Bob labels this point as negative, then $M^- = M^- + 1$ and $W = W+1$.
            \State If Bob labels this point as positive, then Trent sends this point to the court. If this point is a true positive, then $M^+ = M^+ + 1$ and $W = 1$. 
            \State If $M^+ > M^-$, then Trent sends all points in $(-\infty,t_A)$ to Bob.  Bob sends the labels of these points to Trent. Trent sends the points labeled as positive by Bob to the court to settle. 
        \EndFor
        \State Output the labels of documents in $R$. If the document $x$ is not sent to the court, then label $x$ by Alice's report for $x \geq t_A$ and label $x$ as negative for $x < t_A$; otherwise, output the label of this document decided by the court.
    \end{algorithmic}
\end{algorithm}

The Label-Verification protocol for classifier report is guaranteed to achieve $1-3\err^*/N^+$ recall on $(R,f)$ with probability as least $1-\delta$. If the classifier reported by Alice has more than $3\err^*$ classification errors on $(R,f)$, then the protocol detects this classifier reported by Alice is not optimal with probability at least $1-\delta$. 
If the protocol detects this classifier is not optimal, then it reveals all documents in $R$ to Bob for verification. For nice instances with constant optimal error $\err^*$, with a constant failure probability $\delta$, this protocol has $O(\log^2 N)$ non-responsive disclosure if Alice reports the optimal classifier for $(R,f)$. If the number of negative points $N^- \gg N^+$, then the non-responsive disclosure of this protocol is much smaller than the number of all negative points.

The Label-Verification Protocol for Classifier Report is similar to the Label-Verification Protocol for Label Report. The main difference is that the stopping condition for Label-Verification Protocol for Classifier Report (Algorithm~\ref{alg:one_dim_classifier}) is when the classifier $t_A$ reported by Alice is detected to be not optimal instead of detecting one error made by Alice in Label-Verification Protocol for Label Report (Algorithm~\ref{alg:one_dim_label}). 
The classifier reported by Alice is detected to be not optimal if and only if enough positives in $(-\infty,t_A)$ are detected and a better classifier with smaller errors is already found. 
Since the stopping condition is changed, this new protocol might allow multiple errors made by Alice. Therefore, the selection rule for documents revealed to Bob is changed to have multiple epochs accordingly. In each epoch, the document with the $i$-th highest position in this epoch is sampled with probability proportional to $1/i$. Whenever the protocol detects an error made by Alice, the protocol starts a new epoch and resets the sampling probability to $1$. 
In this protocol, Alice is only asked to label points in $[t_A,+\infty)$ not all points. We set $f_A(x)$ to be Alice's report if $x \geq t_A$, and $f_A(x) = -1$ if $x < t_A$. 
Finally, the protocol outputs the label of each document $x$ as the court decision if this document is sent to the court; otherwise, label this document $x$ by $f_A(x)$.
In this output, all documents labeled as positive are correct since they are either verified by Bob or settled by the court.

\subsection{Lower Bound}

In this section, we provide a lower bound on the non-responsive disclosure of protocols with a large recall. 

\begin{restatable}{theorem}{thmlower}
\label{thm:lower-bound}
    For any multi-party binary classification protocol with the recall at least $\REC^*_{(R,f)}$ on all one-dimensional instances $(R,f)$, the worst-case non-responsive disclosure of this protocol is at least $\Omega(\ln N)$ on the instance with $N$ documents,
    where $\REC^*_{(R,f)} = 1-\FN^*_{(R,f)}/N^+_{(R,f)}$ is the recall given by the optimal classifier on $(R,f)$.
\end{restatable}

This lower bound implies that to achieve a recall of at least $\REC^*$ on all one-dimensional instances, the $\log N$ dependence on the number of documents $N$ in non-responsive disclosure is necessary. For instances with constant $\err^*$, with a constant failure probability $\delta$, our Label-Verification Protocol for Label Report has $O(\log N)$ non-responsive disclosure if Alice reports true labels, which matches this worst-case lower bound. For instances with constant $\err^*$ and constant $\delta$, our Label-Verification Protocol for Classifier Report has $O(\log^2 N)$ non-responsive disclosure if Alice reports the optimal classifier.

\begin{proof}[Proof of \Cref{thm:lower-bound}]
    We construct a family of $\log_2 N$ one-dimensional instances as follows. These instances have the same data point positions $R \subseteq \bbR$ with $|R| = N$ data points $x_1 \geq x_2 \geq \cdots \geq x_{N}$. These $\log_2 N$ instances have different labels $f_j$ for $j = 1,\cdots, \log_2 N$. We partition the data points into $\log_2 N$ buckets $B_0 = \{x_1\}$ and $B_j = \{x_i :  2^{j-1}+1 \leq i \leq 2^j \}$ for $j =1, 2,\cdots, \log_2 N$. The label function $f_j$ assigns data points in $B_0 \cup B_j$ as positive points and the rest points as negative points. These $\log_2 N$ instances have different best linear classifiers. The best threshold for the instance $(R,f_j)$ is $t_j^* = x_{i}$ where $i = 2^j$. 
    If the protocol can not distinguish these $\log_2 N$ instances, then the recall of the protocol will be smaller than $1-(\FN^*_{(R,f_j)}+1)/N^+_{(R,f_j)}$ for one of these instances $(R,f_j)$.
    Thus, to distinguish these $\log_2 N$ instances, the protocol needs to choose one point from each bucket. Then, the non-responsive disclosure is at least $\log_2 N$. 
\end{proof}
\section{Can We Have Much Smaller Privacy Loss?} 
\label{sec:crit}

\begin{figure*}[t]
    \centering
    \includegraphics[width=0.45\textwidth]{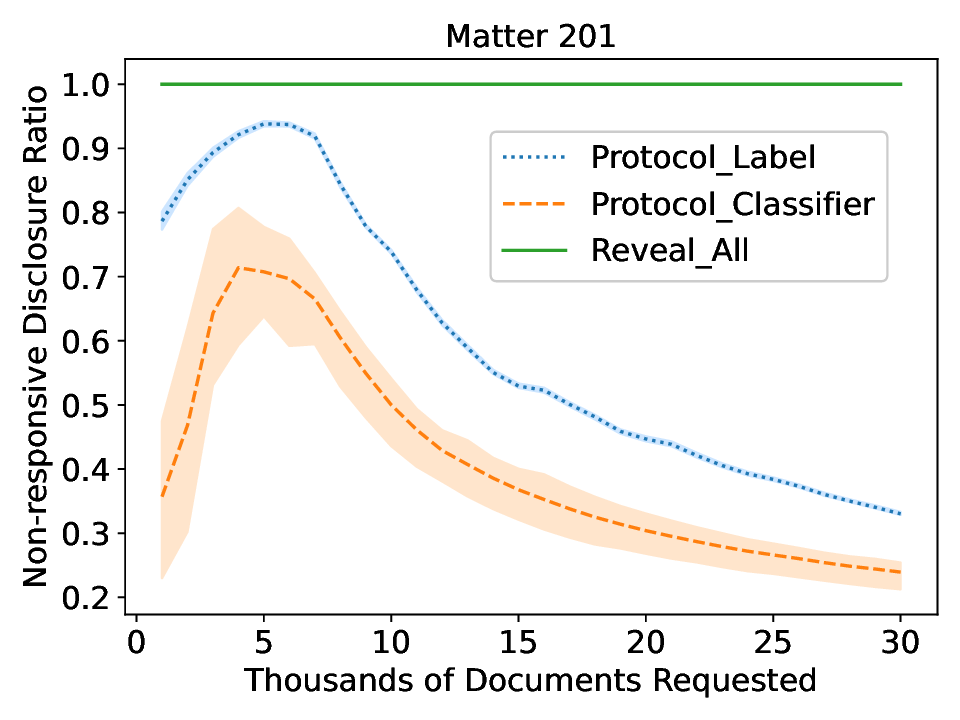}
    \includegraphics[width=0.45\textwidth]{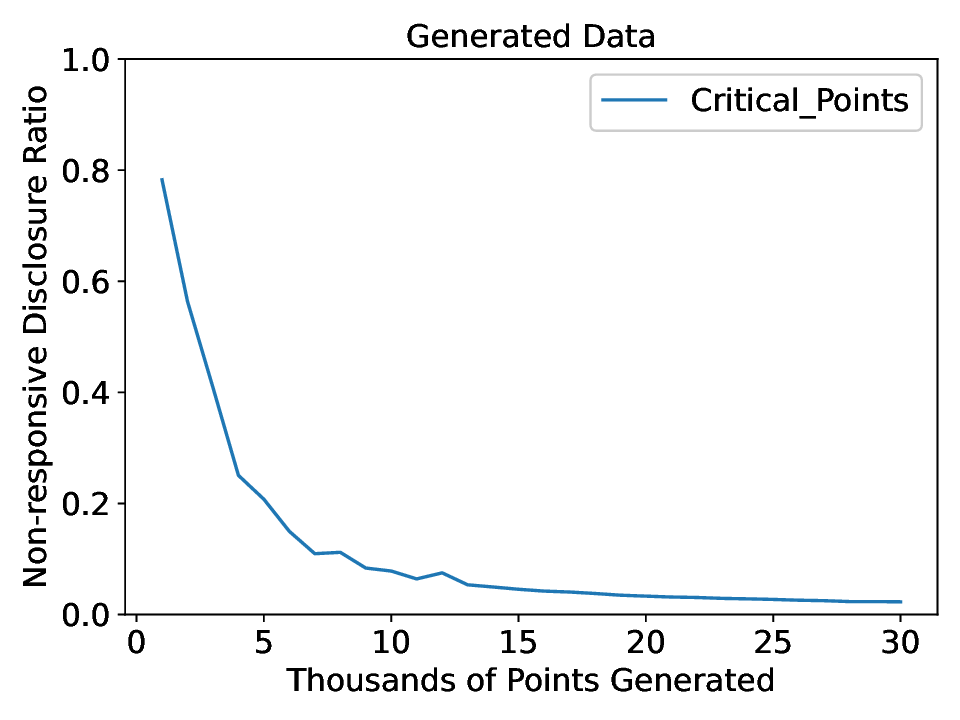}

    \caption{Non-responsive disclosure ratio (the non-responsive disclosure divided by the number of non-responsive documents in the hand-labeled documents) of our protocols on the review task Matter 201 (left panel) and Non-responsive disclosure ratio (the non-responsive disclosure divided by the number of non-responsive points in the generated points) of the critical point protocol on the simulated data (right panel). For the right panel, the data are generated by first sampling from two standard Gaussian and then shifting misclassified points such that the dataset is realizable. The ratio of the number of points in the two types is $0.05$, which is roughly the fraction of responsive documents in Matter 201. \Cref{alg:crit} was used to compute the number of critical points, which is the non-responsive disclosure.}
    \label{fig:crit}
\end{figure*}

In this section, we make an empirical comparison between the non-responsive disclosure of our protocols from \Cref{sec:experiment} with that of the critical points protocol proposed in \citet{dong2022classification}. Note that in the realizable case, the critical points protocol has minimal non-responsive disclosure. Based on this empirical comparison, we conjecture that the non-responsive disclosure of our protocols can not be improved significantly.

We observe in \Cref{fig:crit} that although the non-responsive ratio of the critical points protocol is lower (7\% vs 20\%) than that of our protocol, the asymptotic trend of the non-responsive ratio of our protocols and critical points protocol are quite similar. Note that the critical points protocol has the minimum non-responsive disclosure in the realizable setting. Even in the realizable case, there is a significant and unavoidable amount of non-responsive disclosure. Since the learning problem in the non-realizable setting is much harder than in the realizable setting, we conjecture that the non-responsive disclosure of our protocols is unlikely to have significant room for improvement.

We note that the experiment is more suggestive than rigorous. There are several significant differences between the left and right panels:
\begin{enumerate}
	\item The data distributions are different. The left panel uses real data and is non-realizable, while the dataset used in the right panel is generated by using the mixture of two Gaussians and enforcing the sampled points to be realizable. (The enforcing step shifts the data points that are misclassified by the optimal classifier such that the dataset is realizable. We do not use this enforcing step on the real data used in the left panel because the dimension of this real data exceeds the computational limitation of the critical points protocol.) 
	\item The dimensions of the two datasets are different. The real-world data used in the left panel has dimensions in hundreds of thousands while the generated data in the right panel has only $100$ dimensions. This is more of a computational difficulty: even with our significant improvement, the critical point protocol is still computationally demanding.
\end{enumerate}

In the rest of the section, we overview a fast implementation of the critical points protocol with the details in Appendix~\ref{apx:fast_critical} and hopefully shine new thoughts on further improvements. We first introduce some terms and notations that are useful in this section.
Recall that an instance $(X,f)$ consists of a set of documents $X\subseteq \R^d$ and $f:X\to\{-1,1\}$ indicates the labels of documents. Let $n^+ = |\{x\in X: f(x) = 1\}|$ be the number of responsive documents in $X$ and $n^- = |\{x\in X: f(x)=-1\}|$ be the number of non-responsive documents.

The critical points protocol reveals all positives and critical points in negatives reported by Alice. A document reported as negative by Alice is a critical point if and only if after flipping the label of this document to positive, the labels reported by Alice are still linearly separable. Thus, critical points are those negative points in the boundary of all reported negatives, which serves as a proof for the correctness. 

The algorithm provided in \citet{dong2022classification} requires solving $n$ linear programs, each of size $n$ by $d$. Assuming the best possible solver for the dense linear program (which is not far in reach, see \citet{van2020solving}), this takes $O(n^2d)$ time and is too slow in practice. We provide a fast implementation of critical points protocol by using the algorithm for finding extremal points proposed by~\citet{clarkson1994more}. Specifically, if the number of critical points is $\ell$, then this fast implementation takes $O(nd\ell)$ time, which is much faster than that in~\citet{dong2022classification} since the number of critical points $\ell$ is usually much smaller than $n$. The details of this implementation are discussed in Appendix~\ref{apx:fast_critical}.

\section{Conclusion} 
\label{sec:conclusion_and_future_directions}
In this work, we explore accountability concerns in multi-party classification problems that arise in e-discovery. Our protocol studies the e-discovery protocol setting in \citet{dong2022classification} and gives a new protocol that goes beyond the realizable case. Experiments on real and simulated data show that the three desiderata can be simultaneously achieved: (1) accountability (2) high recall and (3) low non-responsive disclosure. We further prove that high recall is guaranteed even when the defendant (Alice) behaves strategically, but only in the single-dimensional case.

There are many interesting directions for future research. A natural question that is left open by this work is whether we can obtain theoretical guarantees in higher dimensions (analogous to \Cref{thm:one-dim-classifier,thm:one-dim-label}). This may require additional assumptions about the data distribution to overcome the computational intractability of the non-realizable linear classification problem in high dimensions, even without any accountability consideration. Another interesting question is to consider the different levels of privacy significance among documents, which is common in legal practice. It requires to design a more delicate measurement of privacy loss. In this work, we use the non-responsive disclosure as a simplified measurement of privacy loss. 
Finally, it is important to further consider the objectives of the defendant and plaintiff and fully characterize the best response strategy of the defendant (Alice) under natural assumptions.  
It is also interesting to consider different compelling motions when the protocol detects enough errors in Alice's report instead of just revealing all requested documents. 


\bibliographystyle{plainnat}
\bibliography{references}

\begin{thebibliography}{26}
\providecommand{\natexlab}[1]{#1}
\providecommand{\url}[1]{\texttt{#1}}
\expandafter\ifx\csname urlstyle\endcsname\relax
  \providecommand{\doi}[1]{doi: #1}\else
  \providecommand{\doi}{doi: \begingroup \urlstyle{rm}\Url}\fi

\bibitem[Awasthi et~al.(2017)Awasthi, Balcan, and
  Long]{awasthi2017localization}
Pranjal Awasthi, Maria~Florina Balcan, and Philip~M. Long.
\newblock The power of localization for efficiently learning linear separators
  with noise.
\newblock \emph{J. ACM}, 63\penalty0 (6), jan 2017.
\newblock ISSN 0004-5411.
\newblock \doi{10.1145/3006384}.
\newblock URL \url{https://doi.org/10.1145/3006384}.

\bibitem[Chang et~al.(2020)Chang, Yu, Chang, Yang, and Kumar]{chang2020pre}
Wei-Cheng Chang, Felix~X Yu, Yin-Wen Chang, Yiming Yang, and Sanjiv Kumar.
\newblock Pre-training tasks for embedding-based large-scale retrieval.
\newblock \emph{arXiv preprint arXiv:2002.03932}, 2020.

\bibitem[Clarkson(1994)]{clarkson1994more}
Kenneth~L Clarkson.
\newblock More output-sensitive geometric algorithms.
\newblock In \emph{Proceedings 35th Annual Symposium on Foundations of Computer
  Science}, pages 695--702. IEEE, 1994.

\bibitem[Cormack and Grossman(2014)]{cormack2014evaluation}
Gordon~V Cormack and Maura~R Grossman.
\newblock Evaluation of machine-learning protocols for technology-assisted
  review in electronic discovery.
\newblock In \emph{Proceedings of the 37th international ACM SIGIR conference
  on Research \& development in information retrieval}, pages 153--162, 2014.

\bibitem[Cormack and Grossman(2017)]{cormack2017technology}
Gordon~V Cormack and Maura~R Grossman.
\newblock Technology-assisted review in empirical medicine: Waterloo
  participation in clef ehealth 2017.
\newblock \emph{CLEF (working notes)}, 11, 2017.

\bibitem[Cormack and Mojdeh(2009)]{cormack2009machine}
Gordon~V Cormack and Mona Mojdeh.
\newblock Machine learning for information retrieval: Trec 2009 web, relevance
  feedback and legal tracks.
\newblock In \emph{TREC}, 2009.

\bibitem[Dong et~al.(2022)Dong, Hartline, and
  Vijayaraghavan]{dong2022classification}
Jinshuo Dong, Jason Hartline, and Aravindan Vijayaraghavan.
\newblock Classification protocols with minimal disclosure.
\newblock In \emph{Proceedings of the 2022 Symposium on Computer Science and
  Law}, pages 67--76, 2022.

\bibitem[\emph{Brown}~v.\ {\emph{Tellermate Holdings
  Ltd.}}(2014)]{brownvtellermate2014}
\emph{Brown}~v.\ {\emph{Tellermate Holdings Ltd.}}
\newblock \emph{Case No. 2:11-cv-1122, 2014 U.S. Dist. LEXIS 90123}, 2014.
\newblock URL \url{https://casetext.com/case/brown-v-tellermate-holdings-ltd}.

\bibitem[\emph{Hyles}~v.\ {\emph{New York City}}(2016)]{hylesvnyc2016}
\emph{Hyles}~v.\ {\emph{New York City}}.
\newblock \emph{10 Civ. 3119 (AT)(AJP)}, 2016.
\newblock URL \url{https://casetext.com/case/hyles-v-nyc}.

\bibitem[\emph{Moore}~v.\ {\emph{Groupe}}(2012)]{moorevgroupe2012}
\emph{Moore}~v.\ {\emph{Groupe}}.
\newblock \emph{868 F. Supp. 2d 137}, 2012.
\newblock URL \url{https://casetext.com/case/moore-v-groupe}.

\bibitem[Goldreich et~al.(1987)Goldreich, Micali, and
  Wigderson]{goldreich1987play}
O~Goldreich, S~Micali, and A~Wigderson.
\newblock How to play any mental game.
\newblock In \emph{Proceedings of the nineteenth annual ACM symposium on Theory
  of computing}, pages 218--229, 1987.

\bibitem[Goldwasser et~al.(1989)Goldwasser, Micali, and
  Rackoff]{goldwasser1989knowledge}
Shafi Goldwasser, Silvio Micali, and Charles Rackoff.
\newblock The knowledge complexity of interactive proof systems.
\newblock \emph{SIAM Journal on Computing}, 18\penalty0 (1):\penalty0 186--208,
  1989.

\bibitem[Goldwasser et~al.(2021)Goldwasser, Rothblum, Shafer, and
  Yehudayoff]{goldwasser2021pacverification}
Shafi Goldwasser, Guy~N. Rothblum, Jonathan Shafer, and Amir Yehudayoff.
\newblock {Interactive Proofs for Verifying Machine Learning}.
\newblock In James~R. Lee, editor, \emph{12th Innovations in Theoretical
  Computer Science Conference (ITCS 2021)}, volume 185 of \emph{Leibniz
  International Proceedings in Informatics (LIPIcs)}, pages 41:1--41:19,
  Dagstuhl, Germany, 2021. Schloss Dagstuhl--Leibniz-Zentrum f{\"u}r
  Informatik.
\newblock ISBN 978-3-95977-177-1.
\newblock \doi{10.4230/LIPIcs.ITCS.2021.41}.
\newblock URL \url{https://drops.dagstuhl.de/opus/volltexte/2021/13580}.

\bibitem[Grossman and Cormack(2010)]{grossman2010technology}
Maura~R Grossman and Gordon~V Cormack.
\newblock Technology-assisted review in e-discovery can be more effective and
  more efficient than exhaustive manual review.
\newblock \emph{Rich. JL \& Tech.}, 17:\penalty0 1, 2010.

\bibitem[Grossman and Cormak(2012)]{grossman2012inconsistent}
Maura~R Grossman and Gordon~V Cormak.
\newblock Inconsistent responsiveness determination in document review:
  Difference of opinion or human error.
\newblock \emph{Pace L. Rev.}, 32:\penalty0 267, 2012.

\bibitem[Guruswami and Raghavendra(2009)]{guruswami2009hardness}
Venkatesan Guruswami and Prasad Raghavendra.
\newblock Hardness of learning halfspaces with noise.
\newblock \emph{SIAM Journal on Computing}, 39\penalty0 (2):\penalty0 742--765,
  2009.
\newblock \doi{10.1137/070685798}.
\newblock URL \url{https://doi.org/10.1137/070685798}.

\bibitem[Hedin et~al.(2009)Hedin, Tomlinson, Baron, and
  Oard]{hedin2009overview}
Bruce Hedin, Stephen Tomlinson, Jason~R Baron, and Douglas~W Oard.
\newblock Overview of the trec 2009 legal track.
\newblock In \emph{TREC}, 2009.

\bibitem[Kalai et~al.(2008)Kalai, Klivans, Mansour, and
  Servedio]{kalai2008agnostically}
Adam~Tauman Kalai, Adam~R. Klivans, Yishay Mansour, and Rocco~A. Servedio.
\newblock Agnostically learning halfspaces.
\newblock \emph{SIAM Journal on Computing}, 37\penalty0 (6):\penalty0
  1777--1805, 2008.
\newblock \doi{10.1137/060649057}.
\newblock URL \url{https://doi.org/10.1137/060649057}.

\bibitem[Kearns and Vazirani(1994)]{kearnsvazirani1994book}
Michael~J. Kearns and Umesh~V. Vazirani.
\newblock \emph{An Introduction to Computational Learning Theory}.
\newblock MIT Press, Cambridge, MA, USA, 1994.
\newblock ISBN 0262111934.

\bibitem[Kearns et~al.(1992)Kearns, Schapire, and Sellie]{kearns1992toward}
Michael~J. Kearns, Robert~E. Schapire, and Linda~M. Sellie.
\newblock Toward efficient agnostic learning.
\newblock In \emph{Proceedings of the Fifth Annual Workshop on Computational
  Learning Theory}, COLT '92, page 341–352, New York, NY, USA, 1992.
  Association for Computing Machinery.
\newblock ISBN 089791497X.
\newblock \doi{10.1145/130385.130424}.
\newblock URL \url{https://doi.org/10.1145/130385.130424}.

\bibitem[Kluttz and Mulligan(2019)]{kluttz2019automated}
Daniel~N Kluttz and Deirdre~K Mulligan.
\newblock Automated decision support technologies and the legal profession.
\newblock \emph{Berkeley Technology Law Journal}, 34\penalty0 (3):\penalty0
  853--890, 2019.

\bibitem[Louis and Spanakis(2021)]{louis2021statutory}
Antoine Louis and Gerasimos Spanakis.
\newblock A statutory article retrieval dataset in french.
\newblock \emph{arXiv preprint arXiv:2108.11792}, 2021.

\bibitem[O’Mara-Eves et~al.(2015)O’Mara-Eves, Thomas, McNaught, Miwa, and
  Ananiadou]{o2015using}
Alison O’Mara-Eves, James Thomas, John McNaught, Makoto Miwa, and Sophia
  Ananiadou.
\newblock Using text mining for study identification in systematic reviews: a
  systematic review of current approaches.
\newblock \emph{Systematic reviews}, 4\penalty0 (1):\penalty0 1--22, 2015.

\bibitem[Stuart(2021)]{stuart2021right}
Allyson~Haynes Stuart.
\newblock A right to privacy for modern discovery.
\newblock \emph{Geo. Mason L. Rev.}, 29:\penalty0 675, 2021.

\bibitem[van~den Brand et~al.(2020)van~den Brand, Lee, Sidford, and
  Song]{van2020solving}
Jan van~den Brand, Yin~Tat Lee, Aaron Sidford, and Zhao Song.
\newblock Solving tall dense linear programs in nearly linear time.
\newblock In \emph{Proceedings of the 52nd Annual ACM SIGACT Symposium on
  Theory of Computing}, pages 775--788, 2020.

\bibitem[Zou and Kanoulas(2020)]{zou2020towards}
Jie Zou and Evangelos Kanoulas.
\newblock Towards question-based high-recall information retrieval: Locating
  the last few relevant documents for technology-assisted reviews.
\newblock \emph{ACM Transactions on Information Systems (TOIS)}, 38\penalty0
  (3):\penalty0 1--35, 2020.

\end{thebibliography}

\newpage

\appendix

\section{Omitted Proofs}
\label{sec:appendix}

\subsection{Proofs in Section~\ref{sec:label_report}} 
\label{sub:proofs_in_sec:label_report}


In this section, we provide the proofs in the section~\ref{sec:label_report}. To prove Theorem~\ref{thm:one-dim-label}, we first show the following lemma.

\begin{restatable}{lemma}{lemlabel}
\label{lem:one-dim-label}
Given the one-dimensional instance $(R,f)$, error tolerance $k \geq 1$ and failure probability $\delta \in (0,1)$, Label-Verification Protocol for Label Report (Algorithm~\ref{alg:one_dim_label}) detects a true positive point labeled as negative by Alice with probability at least $1-\delta$ if the optimal classifier for Alice’s report
has the true error $\err(t_A^*) \geq \err^*+k$. If Alice reports true labels, the expected number of sampled negatives is at most 
$$
\left(2+\frac{2\,\err_A(t_A^*)}{k}\right) \ln N^- \ln(1/\delta).
$$
\end{restatable}

\begin{proof}[Proof of \Cref{lem:one-dim-label}]
Since Algorithm~\ref{alg:one_dim_label} sends all points reported as positives by Alice to Bob, Alice will not label any negative points as positives. Hence, the optimal threshold $t^*_A$ for Alice's report is at least the optimal threshold $t^*$ for true labels.  Otherwise, suppose $t^*_A<t^*$. Since $t^*$ is the optimal threshold for the true label, we have $\err(t^*_A) - \err(t^*) \geq 0$. Since Alice will not report any negatives as positives, we have $\err_A(t^*_A)-\err_A(t^*) \geq \err(t^*_A)-\err(t^*) \geq 0$, which contradicts that $t^*_A$ is the largest threshold minimizes the error of Alice's report.

Suppose $t^*_A> t^*$ and there are points in $[t^*,t_a)$. We consider all positive points in $[t^*, t^*_A)$ that are labeled as negatives by Alice. We use $I= \{i: x_i \in [t^*, t^*_A), f(x_i) = 1, f_A(x_i) = -1\}$ to denote the indices of these hidden positive points in the sampling sequence of Algorithm~\ref{alg:one_dim_label}. Let $Y_i$ be the indicator variable that the point $x_i$ is sampled by the algorithm. Then, the probability that the algorithm samples no point from $I$ is  
$$
\Pr\left\{\sum_{i \in I} Y_i = 0\right\} = \prod_{i\in I} (1-p_i),
$$
where  $p_i$ is the sampling probability for each point $x_i$ in the sampling protocol.

If the sampling probability $p_i = 1$ for some $i\in I$, then the algorithm always succeeds. We next consider $p_i < 1$ for all $i \in I$. Let $b = |\{i: x_i \in [t^*, t^*_A), f(x_i) = -1\}|$ be the number of true negatives in $[t^*, t^*_A)$. Let $a =|I|$ be the number of positives in $[t^*, t^*_A)$ that are reported as negatives by Alice. We sort indices in $I$ in increasing order $i_1 < i_2 < \cdots < i_a$. For every $i_j \in I$, the negatives reported by Alice before $i_j$ are either true negatives or hidden positives in $\{i_1,\cdots,i_{j-1}\}$, which means $i_j \leq b+j$.
Then, we have for every $i_j \in I$, the point $x_{i_j}$ is sampled with probability at least 
$$
p_{i_j} = \frac{c}{i_j} \geq \frac{c}{b+j}, 
$$
where the second inequality is due to $i_j \leq b+j$. 

Then, the probability that the algorithm samples no point from $I$ is at most
\begin{align*}
\Pr\left\{\sum_{i \in I} Y_i = 0\right\} &= \prod_{i\in I} (1-p_i) \leq \prod_{i\in I} e^{-p_i} \leq \\
&\leq \exp\left(-\sum_{j=1}^a \frac{c}{b+j}\right) \leq \exp\left(-c \ln \frac{b+a}{b}\right).
\end{align*}
Since $\ln(1+x) \geq x/(1+x)$, we have
$$
\left(\ln \frac{b+a}{b}\right)^{-1} \leq \frac{a+b}{a} = 1+\frac{b}{a}. 
$$
Let $a' = |\{i: x_i \in [t^*, t^*_A), f(x_i) = 1\}|$ be the number of true positives in $[t^*, t^*_A)$. If $\err(t^*_A) > \err(t^*) + k$, then we have $a' - b > k$. Since $t^*_A$ is the optimal threshold for Alice's labels, we have $\err_A(t^*_A) < \err_A(t^*)$. Note that $\err_A(t^*)- \err_A(t^*_A) = (b+a) - (a'-a) > 0$. Thus, we have $2a > a'-b > k$, which implies $a>k/2$. Therefore, we have
$$
\left(\ln \frac{b+a}{b}\right)^{-1} \leq 1+\frac{b}{a} \leq 2+\frac{a'-a}{a} \leq 2+\frac{2\,\err_A(t^*_A)}{k},
$$
where the second inequality is due to $b < a'-k < a'$ and the third inequality is from $\err_A(t_A^*) \geq a'-a$ and $a > k/2$.
By taking parameter $c = (2+2\,\err_A(t^*_A)/k) \ln \delta$ in Algorithm~\ref{alg:one_dim_label}, we have 
$$
\Pr\left\{\sum_{i \in I} Y_i = 0\right\} \leq \exp\left(-c \ln \frac{b+a}{b}\right) \leq \delta. 
$$

Finally, if Alice reports true labels, the expected number of sampled points is at most 
$$
\E\left[\sum_{i=1}^m Y_i\right] = \sum_{i=1}^m p_i \leq c \sum_{i=1}^m \frac{1}{i} \leq c \ln m \leq c \ln N^-,
$$
where $m$ is the number of points reported as negatives in $(-\infty,t_A^*)$ which is at most $N^-$.
\end{proof}

\thmlabel*
\begin{proof}[Proof of Theorem~\ref{thm:one-dim-label}]
    We first consider the recall of our protocol.  
    Note that Alice will not report negatives as positives since all documents reported as positives are sent to Bob. Since all documents in $[t_A^*, +\infty)$ are sent to Bob, Alice will not report any positives in $[t_A^*, +\infty)$ as negatives. 
    If $\err(t^*_A) < \err^* + k$, Alice can hide at most $\err^*+k-1$ positives as negatives. Thus, the recall of our protocol is always at least $1- (\err^* + k-1)/n^+$.
    If $\err(t^*_A) \geq \err(t^*) + k$, by Lemma~\ref{lem:one-dim-label},  the protocol will detect a true positive point labeled as negative by Alice with probability at least $1-\delta$.  Thus, with probability $1-\delta$, the protocol will send all points to Bob and the recall is $1$ in this case.

    We now consider the non-responsive disclosure. If Alice reports the true labels $f$, then there is no hidden positive point. Thus, the non-responsive disclosure of our protocol is the number of points sampled by the protocol and the number of true negatives with scores in $[t^*,+\infty)$. The number of true negatives in $[t^*,+\infty)$ is at most $\err^*$. If Alice reports the true labels, we have $\err_A(t_A^*) = \err^*$. Thus, by combining with the bound on the expected number of sampled points in Lemma~\ref{lem:one-dim-label} and $\err_A(t_A^*) = \err^*$, we get the bound on the expected non-responsive disclosure.
\end{proof}

\corlabel*

\begin{proof}[Proof of Theorem~\ref{cor:one-dim-label}] Suppose the defendant reports all labels correctly. Then, the recall is always $1$. By Theorem~\ref{thm:one-dim-label}, the expected non-responsive disclosure is at most $$(2+2\,\err^*/k)\ln N^- \ln(1/\delta) + \err^*.$$ 
Let $B'$ be the outcome of our protocol with the best response of the defendant. Since our protocol is randomized, $B'$ is a random subset $X$. 
Thus, the expected loss of the defendant with the best response is at most the expected loss of reporting all true labels, which is at most
$$
\E[L_A(B')] \leq (2+2\,\err^*/k)\ln N^- \ln(1/\delta)+\err^*.
$$

Now, consider any defendant's report such that the best classifier $t^*_A$ has true error  $\err(t^*_A) \geq \err^* + k$. Let $B$ be the outcome of this report. Then, by Lemma~\ref{lem:one-dim-label}, with probability at least $1-\delta$, our protocol will detect a hidden positive point and send all data points to the plaintiff.
In this case, the recall is $1$ and the non-responsive disclosure is the number of negative points $N^- = |\{x: f(x) = -1\}|$. If the protocol does not detect the hidden positive point, then the recall is $\REC(B) \geq 0$. The non-responsive disclosure is non-negative. Thus, we have the expected loss of the defendant is
$$
\E[L_A(B)] \geq (1-\delta)N^- +\delta(-c\cdot N^+) \geq (1-2\delta) N^-,
$$
where the last inequality is due to the condition $N^- > c\cdot N^+$.

Since $\delta < 1/3$ and $N^- > 3(2+2\,\err^*/k)\ln N^- \ln(1/\delta)+3 \err^*$, we have 
$$
\E[L_A(B)] \geq \frac{N^-}{3}  > (2+2\,\err^*/k)\ln N^- \ln(1/\delta)+\err^* \geq \E[L_A(B')],
$$
which contradicts that $B'$ is the outcome of the best response.

Thus, the best response satisfies that $\err(t^*_A) < \err^* + k$. This implies the recall is always at least $1-(\err^* + k-1)/N^+$. The expected non-responsive disclosure is at most 
\begin{align*}
\E[\NRD(B')] &= \E[L_A(B')] + cN^+\cdot \E[1-\REC(B')] \\
&\leq (2+2\,\err^*/k)\ln N^- \ln(1/\delta)+(c+1)\err^*+ck.
\end{align*}

For $k=1$, since the best response satisfies $\err(t^*_A) < \err^* + 1$, the optimal threshold on the report of the defendant $t^*_A$ is also the optimal threshold for the true labels. This implies that under the best response, the defendant truthfully reports the labels of all points classified as positive by the optimal classifier on the true labels.
\end{proof}

\subsection{Proofs in Section~\ref{sec:classifier_report}} 
\label{sub:proofs_in_sec:classifier_report}

In this section, we prove Theorem~\ref{thm:one-dim-classifier}. First, we show the following lemma.

\begin{restatable}{lemma}{lemclassifier}
\label{lem:one-dim-classifier}
Label-Verification Protocol for Classifier Report (Algorithm~\ref{alg:one_dim_classifier}) detects the classifier reported by Alice is not optimal with probability at least $1-\delta$ if Alice's classifier $t_A$ has the true error $\err(t_A) > 3\,\err(t^*)$.  The expected number of sampled negatives is at most $2\,\err(t_A)\ln N \ln(N/\delta)$.
\end{restatable}
\begin{proof}[Proof of \Cref{lem:one-dim-classifier}]
    If Alice's report $t_A < t^*$, then Trent will send all positive points in $(t^*, \infty)$ to Bob. By checking the labels of points in $[t_A,t^*)$, we can easily detect the classifier reported by Alice is not the optimal classifier.
    
    Now, suppose that $t_A > t^*$. 
    We sort all points in $[t^*,t_A)$ in decreasing order as Algorithm~\ref{alg:one_dim_classifier}, $t_A > x_1 > x_2 > \cdots > x_{m'} > t^*$. 
    We divide this sequence of points into multiple consecutive groups as follows. 
    We find a subset of indices $1 \leq i_1 < i_2 < \cdots < i_q \leq m'$ such that $i_j$ is the first index where the number of true positives is at least the number of true negatives by counting from the index $i_{j-1}+1$ (we set $i_0 = 0$). This means for any $j = 1,2,\cdots, q$
    \begin{align*}
    |\{i: i_{j-1}+1 \leq  i \leq i_j , f(x_i) &= 1\}| \\
    &\geq  |\{i: i_{j-1}+1 \leq  i \leq i_j , f(x_i) = -1\}|, 
    \end{align*}
    and for any $i_{j-1}+1 \leq i' < i_j$
    \begin{align*}
    |\{i: i_{j-1}+1 \leq  i \leq i' , f(x_i) &= 1\}| \\
    &<  |\{i: i_{j-1}+1 \leq  i \leq i' , f(x_i) = -1\}|.
    \end{align*}
    Then, we partition this sequence into $q$ groups according to these indices $i_1 < i_2 < \cdots < i_q$. The group $j$ contains all points in $G_j = \{x_i : i_{j-1}+1 \leq i \leq i_j\}$.

    First, we show that Algorithm~\ref{alg:one_dim_classifier} will sample at least $q$ true positives among points in $[t^*, t_A)$ with probability at least $1-\delta$. We prove this by induction. We show that the algorithm will sample at least $j$ positives from the first $j$ groups with probability at least $1-j\delta/n$. Consider the first group $G_1$. Note that each group contains either a single positive point or the same number of positives and negatives. If $G_1$ contains a single positive point, then this point is sampled with probability $1$. Suppose $G_1$ contains the same number of positives and negatives. Let $I_1 = \{i : f(x_i) = 1\} \cap G_1$ be the indices of positives in $G_1$. Let $m_1 = |I_1|$ be the number of positives (negatives) in $G_1$. By the similar analysis in Lemma~\ref{lem:one-dim-label}, the probability that the algorithm samples no positive point in group $G_1$ is at most 
    $$
    \prod_{i\in I_1} (1-p_i) \leq \prod_{i\in I_1} e^{-p_i} \leq \exp\left(-\sum_{j=1}^{m_1} \frac{c}{m_1+j}\right) \leq \frac{\delta}{N}.
    $$
    Suppose the induction hypothesis holds for the first $j$ groups. Now, we show it also holds for $j+1$ groups. We consider all points after the last sampled positive point in the first $j$ groups to the end of the group $j+1$. The number of positives is strictly larger than the number of negatives in the range. Thus, the algorithm will sample a new positive in this range with probability at least $1-\delta/N$. By the union bound, the algorithm will sample at least $j+1$ positives with probability at least $1-(j+1)\delta/N$.
    
    We then lower bound the number of groups $q$. Let $a = |\{i: s(x_i) \in [t^*,t_A), f(x_i) = 1\}|$ be the number of positives with scores in $[t^*,t_A)$. Let $b = |\{i: s(x_i) \in [t^*,t_A), f(x_i) = -1\}|$ be the number of negatives in $[t^*,t_A)$. We show that the number of groups is at least $a-b$.
    Note that each group contains either a single positive point or the same number of positives and negatives. If $i_q < m'$, then the number of negatives is greater than the number of positives in the remaining points $\{x_i:i_q+1 \leq i \leq m'\}$. Thus, we must have at least $a-b$ groups containing a single positive point.

    Note that the optimal error is at least $\err(t^*) \geq b$. If Alice's classifier $t_A$ has the error $\err(t_A) > 3\err(t^*)$, then we have
    $$
    a-b = \err(t_A)-\err(t^*) > 2\,\err(t^*) > 2b.
    $$
    Thus, the algorithm will sample at least $q \geq a-b >2b$ positives in $[t^*,t_A)$ with probability at least $1-\delta$. In this case, if the algorithm does not stop and keeps sampling till $t^*$, then the number of negatives counted in $[t^*,t_A)$ is at most $a+b - q \leq 2b <q$, which is strictly less than the number of positives sampled.

    Now, we bound the expected privacy loss. We consider the sampling process of the algorithm starting from $W=1$ and before it samples a true positive. We call this an epoch of the algorithm. The expected number of samples in one epoch is at most 
    $$
    \sum_{i=1}^m p_i \leq \sum_{i=1}^m \frac{c}{W} \leq 2\ln(N/\delta) \ln N.
    $$
    We know that the algorithm can sample at most $\err(t_A)$ positives. Thus, the number of epochs in the algorithm is at most $\err(t_A)$.
\end{proof}

\thmclassifier*
\begin{proof}[Proof of Theorem~\ref{thm:one-dim-classifier}]
    We first bound the recall of our protocol. Since our protocol shown in Algorithm~\ref{alg:one_dim_classifier} always sends all data points in $[t_A, +\infty)$ to Bob, the recall of the protocol is at least $1-\err(t_A)/n^+$. By Lemma~\ref{lem:one-dim-classifier}, our protocol guarantees that with probability at least $1-\delta$, the recall is at least $1-3\err^*/N^+$.
    
    We now check the non-responsive disclosure of the protocol. Note that if the defendant (Alice) reports an optimal classifier for the true labels $f$, then the protocol will never get $M^+ > M^-$. Thus, in this case, the non-responsive disclosure is the number of data points sampled by the protocol and the number of true negatives in $[t_A, +\infty)$. By Lemma~\ref{lem:one-dim-classifier}, the expected number of sampled data points is at most $2\,\err^* \ln N \ln(N/\delta)$. Since $t_A < t^*$, the number of true negatives in $[t_A, +\infty)$ is at most $\err^*$.
\end{proof}

\section{Fast Implementation of Critical Points Protocol}
\label{apx:fast_critical}

In this section, we show a fast implementation of the critical points protocol by using the extremal points algorithm proposed by~\citet{clarkson1994more}.

Let $\mathrm{vert} (X) \subseteq X$ be the set of extremal points of the convex hull of $X$. 
Let $X_+$ and $X_-$ be the set of responsive and non-responsive documents, respectively. 
Assume $w\in\R^d, t\in\R$ gives a perfect classifier, i.e. $w\cdot x + b \geq 0$ for all $x\in X_+$ and $w \cdot x + b<0$ for all $x\in X_-$. Assume real numbers $a_1,\ldots,a_d$ and vectors $v_1,\ldots,v_d\in\R^d$ are such that the following $(d+1)\times (d+1)$ matrix is non-singular
\[
\begin{bmatrix}
	b & a_1 &\cdots & a_d\\
	w & v_1 &\cdots & v_d
\end{bmatrix}
\]
Let $P$ be the corresponding fractional linear mapping from $\R^d$ to $\R^d$: for any $x \in X$
\[P(x) = 
\begin{bmatrix}
	 \frac{a_1+\inner{v_1}{x}}{b+\inner{w}{x}} &\cdots & \frac{a_d+\inner{v_d}{x}}{b+\inner{w}{x}}
\end{bmatrix}\]
We will use the traditional set notation $PX=\{P(x): x\in X\}$. Then we have the following characterization of critical points with the extremal points of $PX$.
\begin{restatable}{theorem}{crit}
\label{thm:crit}
The critical points for a linear separable instance $(X,f)$ is 
	$$\mathrm{crit}(X,f) = \{x\in X_-: P(x)\in\mathrm{vert}(PX)\}.$$
\end{restatable}
\begin{algorithm}
\caption{Critical Points}\label{alg:crit}
    \begin{algorithmic}[1]
        \State \textbf{Input:} $X_+,X_-\subseteq \R^d$ that are linear separable, given as $n^+\times d$ and $n^-\times d$ matrices
        \State \textbf{Output:} critical points of $X_+,X_-$.
        \State Run SVM and find the max-margin linear classifier $x\mapsto \inner{w}{x}+b$.
        \State Find a $d+1$ by $d+1$ orthogonal matrix $U$ whose first column is parallel to $
\begin{bmatrix}
    b \\
    w 
\end{bmatrix}$ by QR decomposition
        \State $V=
\begin{bmatrix}
    \mathbf{1}&X_- \\
    \mathbf{1}&X_+ 
\end{bmatrix}\cdot U$
\State Divide each row of $V$ by its first entry, then discard the first column
        \State Compute the extremal points of V, $I=\mathrm{vert}(V)$ (Algorithm~\ref{alg:vert}) 
        
        \State \Return entries in $I$ less than $n^-$
    \end{algorithmic}
\end{algorithm}
\paragraph{Remark} 
\label{par:remark}

	There are two addtional ingredients compared to the theorem, both of which attempt to control the condition number of the matrix we feed into \Cref{alg:vert}:
	\begin{enumerate}
		\item Although any strict separating hyperplane works, we run SVM to find a max-margin classifier. This ensures the first column of $V$ is not too small, since they serve as denominators in line 6.
		\item Although any non-singular matrix works, we compute an orthogonal matrix $U$ in line 4. This avoid introducing extra co-linearity.
	\end{enumerate}
 
\begin{algorithm}[tb]
\caption{Extremal Points \citep{clarkson1994more}}\label{alg:vert}
    \begin{algorithmic}[1]
        \State \textbf{Input:}  the set of documents $X=\{x_1,\dots,x_n\} \subset \bbR^d$
        \State \textbf{Output:} the extremal points of the convex hull of $\{x_1,\ldots,x_n\}$
        \State $I=\{i\}$ where $x_i$ has the largest first coordinate 
        \For{$j = 1,2, \cdots, n$}
            \While {true}
                \State Solve a linear program to find a  hyperplane $v$ separating $x_j$ from $\{x_i:i\in I\}$
                \If{ fail }
                    \State break
                    \Else
                    \State add $j = \argmax_{k \in [n] \setminus I} v \cdot x_k$ to $I$
                \EndIf
            \EndWhile
        \EndFor
        
        \State \Return $I$
    \end{algorithmic}
\end{algorithm}
\section{Label-Verification Protocol for High-dimensional Instance}

In this section, we generalize the sampling protocol to the non-realizable instance in high-dimensional space $\bbR^d$. 

First, we assume access to an oracle that computes all linear classifiers that minimize the error for any non-realizable instance in $\bbR^d$. We also need the following property on the instance.

\begin{definition}
    A classifier $h$ is defined as consistent with an instance if the positive side of $h^*$ contains the intersection of the positive side of all optimal classifiers for this instance.
\end{definition}

We show the following theorem for the multi-party binary classification on high-dimensional instances. We provide our protocol in Algorithm~\ref{alg:high_dim_label}.

\begin{theorem}\label{thm:high-dim-label}
    Consider a document collection instance $(X,f)$ where $X \subseteq \bbR^d$. Suppose the optimal linear classifier for true labels is consistent with the report of the defendant (Alice). Given an oracle that computes all optimal linear classifiers and parameters $\delta \in (0,1)$ and $k \geq 1$, there is a protocol such that 
    the recall is at least $1-(\err^* + k)/n^+$ with probability at least $1-\delta$, where $\err^*$ is the error of the best linear classifier for $(X,f)$.
\end{theorem}

\begin{algorithm}
\caption{High-dimensional Sampling Protocal}\label{alg:high_dim_label}
    \begin{algorithmic}[1]
        \State \textbf{Parameters:} $k \geq 1$ and $\delta \in (0,1)$
        \State Alice sends all points and their labels $\{(x,f_A(x))\}_{x\in X}$ to Trent.
        \State Trent computes all optimal linear classifiers $\{h^*_A\}$ for all points with labels reported by Alice.
        \State Trent sends to Bob all positive points reported by Alice and all points classified as positive by any optimal classifier $h^*_A$. 
        \State Trent chooses a direction vector $v \in \bbR^d$ (Trent can also ask Alice for such a direction). 
        \State Trent goes through all points classified as negative by all optimal classifiers in $\{h^*_A\}$. These negative points are ordered increasingly by their distances to the union of optimal classifier $\{h^*_A\}$ in direction $v$. 
        \For{each negative point $x_i$}
            \State Trent sends this point to Bob with probability $$p_i = \min\{1, c/n_i\}$$, where 
            $$c = \left(2+\frac{2\,\err_A(t^*_A)}{k}\right) \ln(1/\delta).$$
            \State Bob labels this point and sends the label to Trent.
            \State If Bob labels this point as positive, then Trent sends this point to the court. If this point is a true positive, then Trent sends all data points to Bob.  
        \EndFor
    \end{algorithmic}
\end{algorithm}

\begin{lemma}\label{lem:high-dim-label}
Suppose the optimal classifier for true labels is consistent with Alice's report. Given an error tolerance $k \geq 1$, Algorithm~\ref{alg:high_dim_label} detects a true positive point labeled as negative by Alice with probability at least $1-\delta$ if one optimal classifier $h^*_A$ for Alice's report has the true error $\err(h^*_A) \geq \err(h^*) + k$. 
\end{lemma}

\begin{proof}
    Let $h^*$ be the optimal classifier for true labels. Let $A^*=\{h^*_A\}$ be the set of all optimal classifiers for labels reported by Alice. Since the optimal classifier $h^*$ is consistent with Alice's report, the positive side of $h^*$ contains the intersection of the positive side of classifiers in $A^*$. Let $h'$ be the classifier that is parallel to $h^*$ and tangent to the intersection of the positive side of classifiers in $A^*$. 
    
    Then, we consider the following reduction to the one-dimensional case. Let $v$ be the direction of classifier $h^*$. We project all data points in this direction $v$. The data set after projection is in a one-dimensional space, where each data point $x$ is positioned at $\Pi(x) \in \bbR$. Let $\Pi(h^*)$, $\Pi(h')$ denote the projection of two classifiers $h^*$ and $h'$ in the direction $v$. Note that $\Pi(h^*)$, $\Pi(h')$ are two thresholds in $\bbR$. Since $h'$ is tangent to the intersection of the positive side of classifiers in $A^*$, the positive side of $h'$ is contained in the union of the positive side of classifiers in $A^*$. Thus, all data points with value $\Pi(x) > \Pi(h')$ is sent to Bob at Step $4$ of Algorithm~\ref{alg:high_dim_label}. Next, we only need to consider all points between two classifiers $\Pi(h^*)$ and $\Pi(h')$.

    Let $I = \{x : \Pi(h^*) < \Pi(x) < \Pi(h'), f(x) = 1, f_A(x) = -1\}$, which are hidden positive points between two classifiers $\Pi(h^*)$ and $\Pi(h')$. Similar to the analysis for the one-dimensional instance, we still sort points in $I$ in the decreasing order of $\Pi(x)$. Note that this ordering is different from the ordering that we sample these points in Algorithm~\ref{alg:high_dim_label}. 
    Consider each point $x$ in $I$. Let $d(x) = \min_{h^*_A \in A^*}\{d(x, h^*_A)\}$ be the distance of this point to the union of classifiers in $A^*$. We sample this point with probability $p(x) = \min\{1, c/n(x)\}$, where $n(x)$ is the number of negative points with distance to the intersection of the positive side of classifiers in $A^*$ less than $d(x)$. Since the positive side of $h'$ contained the intersection of the positive side of classifiers in $A^*$, we have $n(x)$ is smaller than the number of negative points before $x$ in the order of $\Pi(x)$. Thus, the sampling probability in Algorithm~\ref{alg:high_dim_label} is always at least the sampling probability in Algorithm~\ref{alg:one_dim_label} for the one-dimensional projection. With the same analysis of Lemma~\ref{lem:one-dim-label}, we complete the proof.
\end{proof}

\begin{proof}[Proof of Theorem~\ref{thm:high-dim-label}]
    Note that our protocol guarantees that all points that are classified as positive by any optimal classifier $h^*_A \in A^*$ are sent to the plaintiff (Bob). Thus, the recall of the protocol is at least $1-\err(h^*_A)/n^+$ for an optimal classifier for Alice's report $h^*_A \in A^*$. By Lemma~\ref{lem:high-dim-label}, we get this recall is at least $1-(\err^*+k)/n^+$ where $\err^* = \err(h^*)$ is the error of the best classifier $h^*$ for $(X,f)$. 
\end{proof}

\newpage

\end{document}